\documentclass[twocolumn,twocolappendix]{aastex631}

\newcommand{\msol}{$M_\odot$}

\newcommand{\tq}{t_{\rm Q}}

\newcommand{\lya}{Ly$\alpha$}

\newcommand{\civ}{C~{\small IV}}
\newcommand{\mgii}{Mg~{\small II}}
\newcommand{\oiii}{[O~{\small III}]}
\newcommand{\nii}{[N~{\small II}]}
\newcommand{\feii}{[Fe~{\small II}]}
\newcommand{\sii}{[S~{\small II}]}
\newcommand{\ha}{H$\alpha$}
\newcommand{\hb}{H$\beta$}
\newcommand{\hg}{H$\gamma$}

\usepackage{amsmath,amssymb,amsfonts}
\usepackage[caption=false]{subfig}
\usepackage[capitalise]{cleveref}
\usepackage{layouts}

\begin{document}

\title{BEES: Quasar lifetime measurements from extended rest-optical emission line nebulae at $z\sim6$}

\author[0000-0001-8986-5235]{Dominika {\v D}urov{\v c}{\'i}kov{\'a}}
\affiliation{MIT Kavli Institute for Astrophysics and Space Research, 77 Massachusetts Avenue, Cambridge, 02139, Massachusetts, USA}
\affiliation{Department of Physics, Massachusetts Institute of Technology, 77 Massachusetts Avenue Cambridge, MA 02139}

\author[0000-0003-2895-6218]{Anna-Christina Eilers}
\affiliation{MIT Kavli Institute for Astrophysics and Space Research, 77 Massachusetts Avenue, Cambridge, 02139, Massachusetts, USA}
\affiliation{Department of Physics, Massachusetts Institute of Technology, 77 Massachusetts Avenue Cambridge, MA 02139}

\author[0000-0001-7572-5231]{Yuzo Ishikawa}
\affiliation{MIT Kavli Institute for Astrophysics and Space Research, 77 Massachusetts Avenue, Cambridge, 02139, Massachusetts, USA}

\author[0000-0002-5367-8021]{Minghao Yue}
\affiliation{Steward Observatory, University of Arizona, 933 North Cherry Avenue, Tucson AZ 85718, USA}

\author[0000-0001-9191-9837]{Marianne Vestergaard}
\affiliation{DARK, The Niels Bohr Insitute, University of Copenhagen, Jagtvej 155, DK-2200 Copenhagen N, Denmark}
\affiliation{Steward Observatory, University of Arizona, 933 North Cherry Avenue, Tucson AZ 85718, USA}

\author[0000-0003-0821-3644]{Frederick B. Davies}
\affiliation{Max Planck Institut f\"ur Astronomie, K\"onigstuhl 17, D-69117 Heidelberg, Germany}

\author[0000-0002-4544-8242]{Jan-Torge Schindler}
\affiliation{Hamburger Sternwarte, Universit\"at Hamburg, Gojenbergsweg 112, D-21029 Hamburg, Germany}

\author[0000-0003-3310-0131]{Xiaohui Fan}
\affiliation{Steward Observatory, University of Arizona, 933 North Cherry Avenue, Tucson AZ 85718, USA}

\author[0000-0002-4770-6137]{Fabrizio Arrigoni Battaia}
\affiliation{Max Planck Institut f\"ur Astrophysik, Karl-Schwarzschild-Str. 1, 85748 Garching, Germany}

\author[0000-0002-3216-1322]{Marta Volonteri}
\affiliation{Institut d’Astrophysique de Paris, CNRS, Sorbonne Université, UMR7095, 98bis bd Arago, 75014 Paris, France}

\author[0000-0003-3769-9559]{Robert A.\ Simcoe}
\affiliation{MIT Kavli Institute for Astrophysics and Space Research, 77 Massachusetts Avenue, Cambridge, 02139, Massachusetts, USA}
\affiliation{Department of Physics, Massachusetts Institute of Technology, 77 Massachusetts Avenue Cambridge, MA 02139}

\author[0000-0002-7054-4332]{Joseph F.\ Hennawi}
\affiliation{Leiden Observatory, Leiden University, P.O. Box 9513, 2300 RA Leiden, The Netherlands}
\affiliation{Department of Physics, University of California, Santa Barbara, CA 93106, USA}

\author[0000-0002-2183-1087]{Laura Blecha}
\affiliation{Department of Physics, University of Florida, Gainesville, FL 32611, USA}

\author[0000-0001-6102-9526]{Irham T. Andika}
\affiliation{Technical University of Munich, TUM School of Natural Sciences, Department of Physics, James-Franck-Str. 1, 85748 Garching, Germany}
\affiliation{Max Planck Institut f\"ur Astrophysik, Karl-Schwarzschild-Str. 1, 85748 Garching, Germany}

\author[0000-0001-8582-7012]{Sarah E. I. Bosman}
\affiliation{Institute for Theoretical Physics, Heidelberg University, Philosophenweg 12, D–69120 Heidelberg, Germany}
\affiliation{Max Planck Institut f\"ur Astronomie, K\"onigstuhl 17, D-69117 Heidelberg, Germany}

\author[0000-0002-4554-4488]{Rebekka Bieri}
\affiliation{Department of Astrophysics, University of Zurich, 8057 Zurich, Switzerland}

\correspondingauthor{Dominika {\v D}urov{\v c}{\'i}kov{\'a}}
\email{dominika@mit.edu}



\begin{abstract}
Measurements of quasar lifetimes at high redshift indicate that the earliest billion-solar-mass supermassive black holes (SMBHs) have only been active as luminous quasars for less than a million years.
Recently, extended \lya\ nebulae around $z\sim6$ quasars have revealed that these short observed lifetimes are unlikely a sightline-dependent effect. However, the interpretation of \lya\ emission is not straightforward due to its resonant nature. In this work, we use rest-frame optical emission lines, which more directly trace photoionization by the quasar, to unambiguously validate the short line-of-sight quasar lifetimes observed at early cosmic epochs. We use deep James Webb Space Telescope/NIRSpec IFU observations of five $z\sim 6$ quasars with small proximity zones to search for their extended emission line nebulae in \ha\ and \oiii$5007$, and detect extended emission in both emission lines around four quasars in our sample. We then use the light-crossing time of these nebulae to measure quasar lifetimes along transverse sightlines. Using their \ha\ nebulae, we also confirm that recombination is likely the dominant emission mechanism behind their previously detected \lya\ nebulae. Our results confirm the existence of high-redshift quasars with extremely short lifetimes, $t_{\rm Q} \lesssim 10^{5}\ {\rm yr}$, hosting billion-solar-mass black holes, indicating that rapid accretion is likely responsible for the assembly of SMBHs in the early Universe.
 
\end{abstract}



\section{Introduction} \label{sec:intro}



The challenge of growing supermassive black holes (SMBHs) at high redshift is an important unsolved problem in modern astrophysics. More than $500$ actively accreting billion-solar-mass BHs, hosted in luminous quasars, have been discovered beyond redshift $z\gtrsim 5$ \citep[e.g.][and references therein]{fan_quasars_2023}, and many more are expected to be discovered in the near future through observatories like Vera C. Rubin Observatory, Euclid or the Nancy Grace Roman Space Telescope. At these cosmic epochs, the Universe is $\lesssim 10^9\ {\rm yr}$ old, providing barely enough time to grow a billion-solar-mass SMBH from a stellar-mass seed -- growing these SMBH would likely require massive seeds and a continuous, Eddington-limited accretion throughout the entire Hubble time \citep{inayoshi_assembly_2020}.

However, recent measurements of high-redshift quasar duty cycles -- the fraction of cosmic time that SMBHs are active as quasars -- have shown that quasars do not shine continuously for the duration of the Hubble time. By comparing the clustering and the number density of luminous quasars at $z\sim6$ to their underlying dark matter halos, \cite{eilers_eiger_2024} have shown that only a small fraction of SMBH-hosting halos exhibit quasar activity, i.e. the duty cycle is in fact $f_{\rm duty} \ll 1$. Such low duty cycles are also found when translating the halo mass function in simulations to the observed quasar luminosity function \citep[e.g.][]{bolgar_imprints_2018}. This means that quasars tend to be quite short-lived, on average yielding integrated quasar lifetimes of only $\tq\approx10^{6.5}\ {\rm yr}$, in contrast to the $\sim10^{9}\ {\rm yr}$ that are expected from continuous, Eddington-limited accretion. This poses a serious challenge to our understanding of SMBH growth in the early Universe.

Such short duty cycles are also supported by constraints on the lifetimes of luminous quasars from their proximity zones \citep{eilers_first_2018,davies_constraining_2020,eilers_detecting_2020,eilers_detecting_2021,morey_estimating_2021,yue_detecting_2023,durovcikova_chronicling_2024}. These measurements use the response of the intergalactic medium (IGM) to the quasar's ionizing radiation that is released during accretion to measure the timescale of quasar activity. In fact, proximity zones have revealed a population of quasars that seem to have been active for even less than a million years, with some lifetimes as short as $\tq\lesssim10^{4}\ {\rm yr}$ \citep[corresponding to line-of-sight proximity zone sizes of $R_{\rm p} \lesssim 2~{\rm pMpc}$\footnote{Note that ${\rm pMpc}$ and ${\rm pkpc}$ stand for proper megaparsec and proper kiloparsec throughout this paper.}\footnote{Note that $R_{\rm p}$ is observationally defined as the distance between the quasar rest frame \lya\ emission and the redshift at which the quasar's spectral flux density drops below $10\%$ of the continuum level, after being smoothed by a boxcar window of width $20\ {\rm \AA}$ \citep{fan_constraining_2006}.},][]{eilers_first_2018,eilers_detecting_2020,eilers_detecting_2021,yue_detecting_2023}. Even though these measurements are mostly sensitive to the last UV-luminous\footnote{``UV-luminous'' refers to an unobscured quasar phase, i.e. when the ultraviolet (UV) radiation released by the SMBH accretion can be observed.} episode, particularly when the IGM neutral gas fraction is low \citep{durovcikova_chronicling_2024}, they do show that phases of rapid accretion tend to be very short-lived at high redshifts.

One way to explain such short UV-luminous lifetimes is by invoking obscuration effects \citep{davies_evidence_2019,satyavolu_need_2023}, as both proximity zone measurements and clustering measurements rely on the escape of UV photons from the quasar along our line of sight. If obscuration effects in the past prevented ionizing photons from reaching the IGM along our line of sight, the size of the observed proximity zones would be reduced. Likewise, a large population of obscured, accreting SMBHs, which are not observed as luminous quasars, could explain the extremely short duty cycles from quasar clustering \citep{eilers_eiger_2024}. The plausibility of obscuration effects is also supported by simulations showing that the UV bright phases are short-lived due to an interplay between SMBH feeding and feedback \citep{trebitsch_black_2019}.

Past line-of-sight obscuration can be probed by measuring quasar lifetimes using imprints of quasar's ionizing radiation in the \textit{transverse} direction \citep[for low-redshift measurements see][]{trainor_constraints_2013, cantalupo_cosmic_2014, hennawi_quasar_2015, borisova_constraining_2016}. As quasars photoionize gas clouds in their circumgalactic medium (CGM), the projected size of the resultant extended nebular emission, $d^{\rm neb}$, can be used to constrain their lifetimes via the light crossing time, i.e. $t_{\rm Q}^{\rm neb} = cd^{\rm neb}$ (where $c$ is the speed of light). This method is sensitive to a different physical mechanism (light travel time) than line-of-sight proximity zones (speed of the IGM ionization front), and so constitutes and independent way of measuring quasar lifetimes. Recently, \cite{durovcikova_quasar_2025} have applied this nebular measurement to a sample of $z\sim6$ quasars with small line-of-sight proximity zones, using deep integral field unit (IFU) observations with the Very Large Telescope/Multi-Unit Spectroscopic Explorer \citep[MUSE,][]{bacon_muse_2010}, and found a correlation between the line-of-sight extent of their proximity zones and the transverse extents of their \lya\ nebulae, as well as the corresponding inferred lifetimes. This result suggests that sightline-dependent obscuration effects are unlikely to play a dominant role in explaining the short UV-luminous quasar lifetimes\footnote{In one case, \cite{durovcikova_quasar_2025} did find a discrepancy between the line-of-sight and transverse UV-luminous lifetimes, which can be attributed to an extremely metal-poor absorption system truncating the line-of-sight proximity zone \citep{durovcikova_extremely_2025}. Note that this quasar is also a part of our sample in this work.}.

However, due to its resonant nature, \lya\ line emission might not directly trace the recombining gas -- instead, the emission could potentially arise from multiple scatterings of \lya\ photons \citep{costa_agn-driven_2022}, which would render the lifetimes found by \cite{durovcikova_quasar_2025} lower limits. To make more robust lifetime measurements using the nebular emission in the CGM, rest-frame optical emission lines, such as \ha\ or \oiii$\lambda \lambda$4959,5007, need to be used. The \ha\ line is particularly useful for this exercise as it traces the same gas phase as \lya\ arising from recombination, and can be thus used to constrain the emission mechanism behind the nebulae observed by \cite{durovcikova_quasar_2025} \citep[e.g. as presented by][]{leibler_detection_2018,peng_direct_2025}.

In this work, we leverage deep ($>3\ {\rm hr}$) James Webb Space Telescope (JWST)/NIRSpec IFU observations of $z\sim6$ quasars to measure nebular lifetimes using the rest-frame optical extended emission in their CGM. Our sample consists of quasars that exhibit some of the shortest line-of-sight proximity zones at high redshift, hence expected to have the smallest nebulae.
Specifically, we search for nebular emission in \ha\ and \oiii, the former directly tracing the dense and cold ($n\approx 1\ {\rm cm^{-3}}$, $T\approx 10^4\ {\rm K}$) recombining hydrogen gas and the latter being a useful probe of the warmer phase of the CGM. Moreover, the extended \lya\ emission has been analyzed by \cite{durovcikova_quasar_2025} for four out of the five quasars in our sample, two of which resulted in detections, which allows us to place constraints on the mechanism behind their \lya\ emission.

This paper is focused on constraining quasar lifetimes and the scientific challenge of explaining the observed SMBH growth. We first present the data and describe the data reduction and preprocessing in \S~\ref{sec:data}. In \S~\ref{sec:BHmasses} we present the rest-optical spectra of these quasars for the first time and use the \ha\ and \hb\ emission lines to estimate their BH masses, confirming the presence of billion-solar-mass BHs at $z\sim6$. We then describe our search for extended emission and present the corresponding lifetime measurements in \S~\ref{sec:nebulae}, interpreting our results in \S~\ref{sec:implications}. In a companion paper (Ishikawa et al. 2025, in prep.), we further analyze the kinematics of the extended emission and present spatially-resolved BPT diagrams, which show that the outer parts of these nebulae are predominantly ionized by the quasar instead of star formation from their host galaxies.

Throughout this paper, we use the flat $\Lambda$CDM cosmology with $h = 0.67$, $\Omega_M=0.31$, $\Omega_\Lambda=0.69$ \citep{planck_collaboration_planck_2020}.

\section{Data}\label{sec:data}

\subsection{Sample of quasars with small proximity zones}

We focus on a sample of $z\sim 6$ quasars exhibiting small line-of-sight proximity zones 
with existing deep NIRSpec IFU observations of their rest-frame optical emission. An overview of this sample is provided in \cref{tab:sample}. Four of these quasars were targeted by the \textit{Black hole Extended Emission Search} (BEES) program (PI Eilers, JWST Cycle 2 GO, Program ID 3079). Additionally, we complement this sample with archival deep NIRSpec IFU observations of one quasar, SDSS J0100+2802, from the JWST Cycle 1 GTO Program (ID 1218, PI Luetzgendorf). Note that the BEES sample was observed with the G395M/F290LP grating/filter combination, while SDSS J0100+2802 was observed with the G395H/F290LP setup, providing the same spectral coverage but higher spectral resolution. 

The BEES quasars were observed between June and October 2024 using an 8-point cycling dither pattern. Observations of SDSS J0100+2802 come from November 2023 and used a 6-point cycling dither pattern. In all cases, no dedicated background or {\it leakcal\footnote{\url{https://jwst-docs.stsci.edu/jwst-near-infrared-spectrograph/nirspec-operations/nirspec-ifu-operations/nirspec-msa-leakage-correction-for-ifu-observations}}} observations were taken.

\begin{table*}[t!]
\centering
\caption{The quasar sample used in this study. $M_{1450}$ is the absolute magnitude at the rest-frame wavelength of $1450$ \AA. Note the extremely small line-of-sight proximity zone sizes, $R_{\rm p}$ (in units of proper Mpc, pMpc), and their corresponding short lifetimes, $\log t_{\rm Q}^{R_{\rm p}}$ (in units of years). All uncertainties are 1$\sigma$ errors. Ref. column contains the reference from which the $R_p$ and $t_Q$ measurements are adopted. E21 - \cite{eilers_detecting_2021}; D20 - \cite{davies_constraining_2020}; E18 - \cite{eilers_first_2018}.}\label{tab:sample}
\begin{tabular}{lcccccccc}
\hline\hline
Quasar & R.A. & Dec. & Redshift & $M_{1450}$ & $R_{\rm p}$ & $\log t_Q^{R_{\rm p}}$ & Ref. & Exp. Time \\
& [hh:mm:ss.ss] & [dd:mm:ss.s] & & [mag] & [pMpc] & [yr] & & [s]\\
\hline
SDSS J0100+2802 & 01:00:13.02 & +28:02:25.8 & 6.327 & $-29.14$ & $7.12 \pm 0.13$ & $5.1^{+1.3}_{-0.7}$ & D20 & 22058.400 \\
PSO J158--14 & 10:34:46.51 & $-$14:25:15.9 & 6.0685 & $-27.41$ & $1.95 \pm 0.14$ & $3.8^{+0.4}_{-0.3}$ & E21 & 11787.824\\
SDSS J1335+3533 & 13:35:50.81 & +35:33:15.8 & 5.9012 & $-26.67$ & $0.78 \pm 0.15$ & $3.0 \pm 0.4$ & E18 & 11787.824 \\
CFHQS J2100--1715 & 21:00:54.62 & $-$17:15:22.5 & 6.0806 & $-25.55$ & $0.37 \pm 0.15$ & $2.3 \pm 0.7$ & E21 & 11787.824 \\
CFHQS J2229+1457 & 22:29:01.65 & +14:57:09.0 & 6.1517 & $-24.78$ & $0.47 \pm 0.15$ & $2.9^{+0.8}_{-0.9}$ & E21 & 11787.824 \\
\hline
\end{tabular}
\end{table*}

\subsection{Data reduction}

All NIRSpec IFU data in this sample were processed with the JWST pipeline\footnote{\url{https://doi.org/10.5281/zenodo.7229890}} (version 1.14.1, CRDS version 1240) with additional steps described below. Note that the only difference between data observed with the G395M and the G395H gratings is that the latter high-resolution grating projects onto both NRS1 and NRS2 detectors, while the medium grating projects onto the NRS1 detector only.

During the first stage of the pipeline (\texttt{spec1}), we add a custom implementation of the NSClean package (version 1.9, \citeauthor{rauscher_nsclean_2024} \citeyear{rauscher_nsclean_2024}) to remove vertical striping in the data by the NIRSpec detectors due to $1/f$ noise.

After the second stage of the pipeline (\texttt{spec2}), we perform a custom outlier rejection procedure on the \texttt{\_cal} files to remove the remaining hot pixels and leaks imprinted through the micro-shutter assembly (MSA). We remove outliers in two steps. First, we mask all pixels with surface brightness above a very conservative threshold ($\sim 2000\ {\rm MJy/sr}$) to remove the largest remaining outliers, often due to cosmic rays. Second, we focus on removing MSA leaks using information from the \texttt{\_cal} files corresponding to the different dithers. Here we utilize the fact that the MSA print-through affects the detector in the same way across all (or most) dithers, while stray light leaking through the failed open MSA shutters typically only affects a small number of dithers. Therefore, we inspected the \texttt{spec2} \texttt{\_cal} files and set a more stringent outlier rejection threshold, such that we are safely not removing any flux from the brightest emission line regions of the quasar, and subsequently mask only outliers that either a) affect the same pixels in all or most dithers, or b) only affect the same pixels in a small number of dithers. Additionally, as MSA leaks tend to be narrowband (i.e. narrow in spectral range), we also only mask the outliers that only span less than $5$ consecutive spectral channels.

Lastly, in \texttt{spec3}, which combines the \texttt{\_cal} files from individual dithers into a final data cube, we use the default spatial resolution of the combined data cube by setting the \texttt{scalexy} parameter to $0.1''$.

\subsection{Further data preprocessing}\label{sec:preprocessing}

In order to make the data cubes science-ready, a few additional steps were performed on the combined data cubes output by the JWST pipeline. Note that we are ultimately interested in searching for faint, extended emission around these quasars, and so the following steps were taken to remove artifacts that could contaminate our search.

\begin{itemize}
    \item First, we masked out edge pixels that have high surface brightness uncertainty. To do this, we masked all pixels at the edges of the data cube ($>1.2''$ away from the quasar) whose variance fell into the top $10$th percentile in each spectral channel.
    \item We perform background subtraction as follows. We model the background in each spectral channel directly from the data by computing the average surface brightness of pixels $>1''$ away from the quasar. However, since the field-of-view of the NIRSpec IFU is limited ($3''\times 3''$, corresponding to $\sim 17\ {\rm pkpc}$ at $z=6$) and could easily be filled with extended emission, we mask the spectral regions corresponding to major emission line regions (\ha, \hb $+$\oiii) when computing the background model. As the background is known to be smoothly varying with wavelength, we then smooth over the thus-computed background at each wavelength and interpolate through the masked emission line regions to create our final background model for each quasar. This background model is subsequently subtracted from the data cube (see Appendix \ref{app:BG} for the final background model for each quasar).
    \item The reduced data cubes still contain several broadband artifacts from the MSA leakage that need to be masked as they can contaminate our PSF subtraction and consequently hinder the nebular search. These artifacts affect the same pixels in the \texttt{\_cal} files across all quasars in our sample and across all dither positions. In an attempt to mask them, we collapse the data cube for $\lambda_{\rm obs} > 50000\ {\rm \AA}$ (where the quasar emission is at its faintest and no extended emission is expected) and mask the brightest pixels at a radius $\gtrsim 1''$ away from the quasar. With these artifacts masked, we rerun our background model calculation as these artifacts could have biased its derived wavelength dependence, and subtract the new background model from the masked data cube (note that this is the model displayed in \cref{fig:BG} in Appendix \ref{app:BG}).
    \item Finally, we use the variance extension of the data cube to mask out spectral channels with outlying variance, which is a sign of remaining hot pixels or strong leaks that escaped our outlier rejection. On average, this step only masks $\sim10$ spectral channels per object. The masked channels also tend to be near the edges of the spectral coverage, and so this step does not affect our extended emission search near the major spectral line regions.
\end{itemize}

\section{Black hole mass measurements}\label{sec:BHmasses}

With the science-ready data cubes at hand, we proceed by extracting the 1D spectrum of the nuclear quasar emission. The aims here are twofold. First, this dataset provides the first look at the rest-optical spectra of these high-redshift quasars (with the exception of SDSS J0100+2802 whose \hg\ and \hb$+$\oiii\ spectrum was previously observed using the NIRCam Wide-Field Slitless Spectroscopy mode, \citeauthor{eilers_eiger_2023} \citeyear{eilers_eiger_2023}). This presents an opportunity to measure the black hole masses of these quasars from their rest-optical emission lines (\ha\ and \hb) for the first time. This is particularly relevant in the case of this sample of quasars that have small proximity zones and thus short inferred lifetimes, which pose the greatest challenge to our SMBH growth models.

\subsection{Spectral fitting}

To extract the nuclear spectrum from the data cube, we integrate the surface brightness over a circular aperture centered at the location of the quasar with a radius of $0.2''$ (approximately corresponding to the full width at half maximum, FWHM, of our PSF model, see Ishikawa et al. 2025, in preparation). The location of the quasar is primarily set by its sky coordinates and is further refined by localizing the brightest pixel near this location in each data cube.

Subsequently, we run Markov chain Monte Carlo (MCMC) using the python package \texttt{emcee} \citep{foreman-mackey_emcee_2013}. As a baseline, we fit the spectral region $2.9\ {\rm \mu m} \leq \lambda_{\rm obs} \leq 5.0\ {\rm \mu m}$ ($2.9\ {\rm \mu m} \leq \lambda_{\rm obs} \leq 5.2\ {\rm \mu m}$ for the higher-redshift SDSS J0100+2802) using the following components: a) a power-law component modeling the continuum emission, defined by its amplitude and spectral slope, b) an iron template\footnote{\url{https://doi.org/10.26093/cds/vizier.34170515}} from \cite{veron-cetty_unusual_2004} representing the \ion{Fe}{2} emission complex, modeled here using two, broad and narrow, kinematical components, c) \ha, \hb, and \hg\ emission lines, each modeled using two (a narrow and a broad) Gaussian components, d) \oiii$\lambda \lambda$\,$4959,5007$ doublet modeled as two Gaussian components and freezing the line ratio to be $1$:$3$ \citep{osterbrock_astrophysics_2006}, e) \oiii$\lambda\,4364$ and \nii$\lambda \lambda$\,$6549,6585$ doublet \citep[line ratio fixed to $1$:$3$;][]{dojcinovic_flux_2023}, each modeled using a single Gaussian component only. Each Gaussian emission line component is defined by its amplitude, linewidth, and offset from the systemic redshift. In all cases, we constrain the doublet line components to share these line properties as atomic physics predicts that they arise from the same gas region.

This procedure results in a good fit for one of the objects, CFHQS J2229+1457, which exhibits pronounced narrow emission lines, while we need to implement further constraints to fit the rest of the spectra due to their relatively weak \oiii\ lines that blend with the enhanced \feii\ emission.
In particular, for all other quasars we model the complex line profiles of \ha\ and \hb\ with two broad Gaussian components instead of one, and we also model the \sii$\lambda \lambda$$\,6718,6732$ doublet (with shared linewidths and velocity offsets but without a fixed line ratio) in order to accurately fit the red wing of the broadest \ha\ component. We further tie the linewidths and velocity offsets of the narrow \ha\ component to the single-Gaussian components of the \nii$\lambda \lambda$\,$6549,6585$ and \sii$\lambda \lambda$\,$6718,6732$ emission lines. Similarly, we tie the narrow component of \hb\ to the narrow components of \oiii$\lambda \lambda$\,$4959,5007$. The extracted quasar spectra as well as the resultant spectral fits are displayed in \cref{fig:spectral_fits}.

\begin{figure*}
    \centering
    \includegraphics[width=\linewidth,trim={0 0.2cm 0 0.1cm},clip]{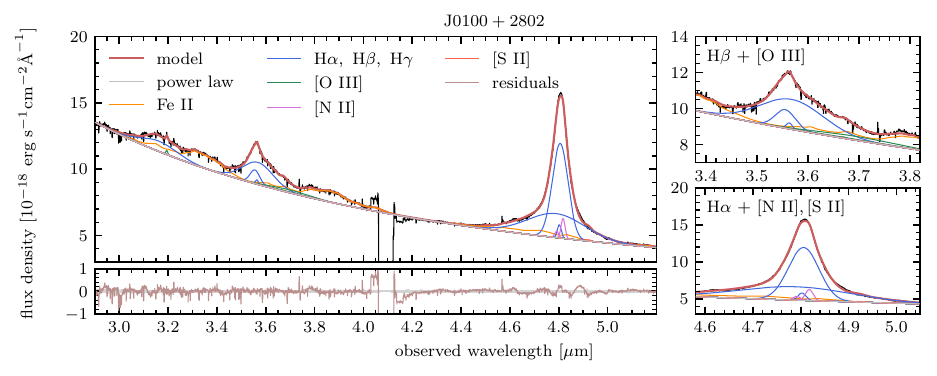}
    \includegraphics[width=\linewidth,trim={0 0.2cm 0 0.1cm},clip]{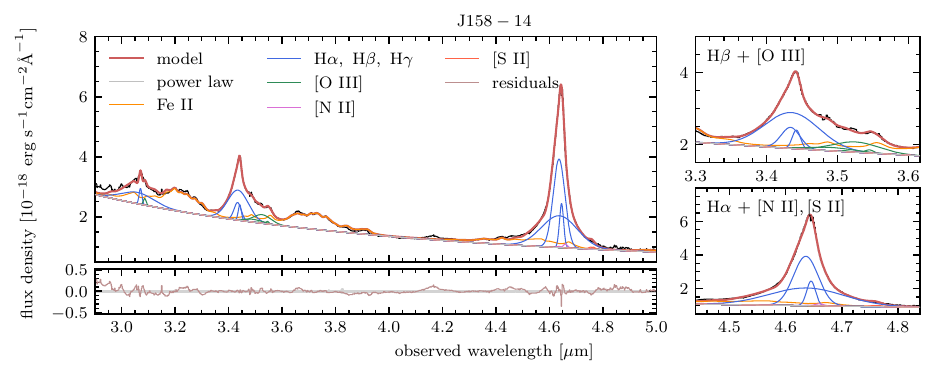}
    \includegraphics[width=\linewidth,trim={0 0.2cm 0 0.1cm},clip]{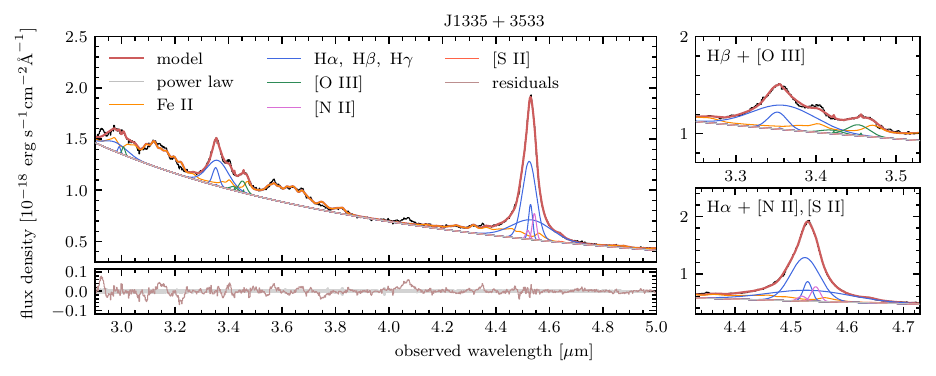}
    \caption{Rest-optical spectra of quasars in our sample extracted over a $0.2"$ (radius) aperture from the NIRSpec IFU observations presented here. For each quasar, we display the full spectrum (top left), the fit residuals (bottom left, pink), as well as a zoom-in on the \hb\ and \ha\ emission line regions (top and bottom right, respectively). The various colored lines represent the median of the MCMC posterior distribution for the different spectral components, with the composite model shown in red. Note that the $\pm1\sigma$ measurement uncertainties on the extracted spectra are shown as gray shaded regions in the residual panel.}
    \label{fig:spectral_fits}
\end{figure*}

\begin{figure*}\ContinuedFloat
    \centering
    \includegraphics[width=\linewidth,trim={0 0.2cm 0 0.1cm},clip]{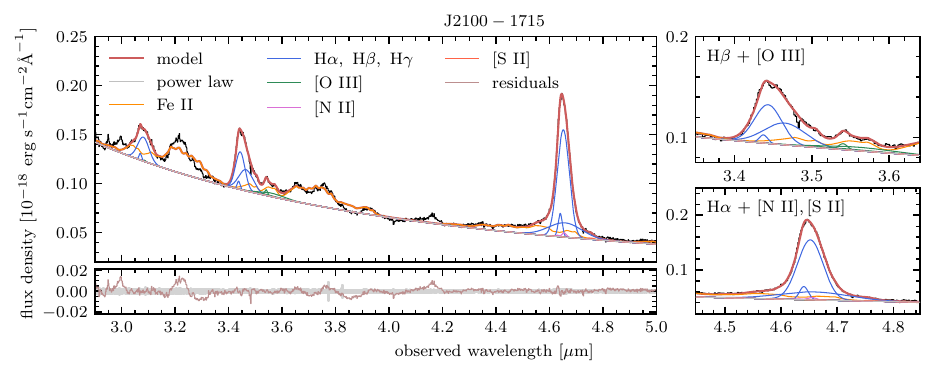}
    \includegraphics[width=\linewidth,trim={0 0.2cm 0 0.1cm},clip]{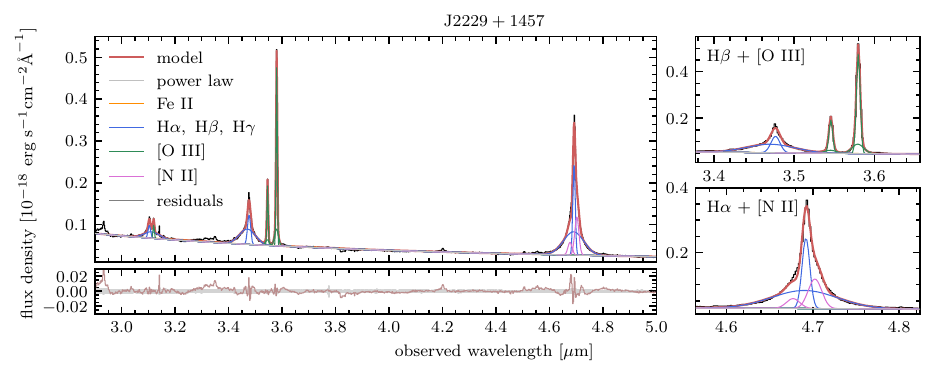}
    \caption{(contd.)}
\end{figure*}

\subsection{Black hole mass indicators}

In high-redshift quasars, black hole masses have been widely estimated using the single-epoch virial method \citep[e.g.][]{shen_comparing_2012}. This method assumes that the broad-line region (BLR) gas clouds undergo virial motion around the central black hole at a distance given by the empirical correlation between the size of the BLR and the quasar luminosity derived from reverberation mapping. Before the advent of JWST, the measurement of black hole masses at high redshift has been limited to scaling relations using rest-frame UV emission line indicators, like \mgii\ or \civ\ \citep[e.g.][]{vestergaard_determining_2006,vestergaard_mass_2009}. However, these are only indirectly calibrated to BH mass measurements from reverberation mapping at low redshift, which primarily use the \hb\ line.

With the access to the rest-frame optical emission lines, we can now use the \ha\ and \hb\ emission lines to independently estimate the BH masses for the quasars in this sample, and thus cross-check the existing rest-UV measurements from the literature. In what follows, we summarize the different methods we use to estimate BH masses from the rest-optical spectra of our quasar sample.

\subsubsection{\hb}

The linewidth of the broad \hb\ line has been considered the most reliable single-epoch estimator. Here, we measure the full width at half maximum (FWHM) of the broad \hb\ component(s) and employ the scaling relation of \cite{vestergaard_determining_2006},
\begin{equation}\label{eq:hb}
    \frac{M_{\rm BH}}{M_{\odot}} = 10^a \left( \frac{{\rm FWHM}_{\rm H\beta}}{1000\ {\rm km\ s}^{-1}} \right)^b \left( \frac{\lambda L_{\lambda}}{10^{44}\ {\rm erg\  s}^{-1}} \right)^c,
\end{equation}
with $a=6.91$, $b=2$, and $c=0.5$. This relation has a systematic scatter of $\sim 0.5\ {\rm dex}$; taking into account the uncertainty in the reverberation mapping measurements, the absolute uncertainty on the BH masses can be as high as $\sim 0.6\ {\rm dex}$ \citep{vestergaard_determining_2006}. Note that this relation has been used to derive BH masses from JWST spectra in existing high-redshift quasar studies \citep[e.g.][]{yang_spectroscopic_2023,eilers_eiger_2023,yue_eiger_2024}.

\subsubsection{\ha}

Alternatively, the broad \ha\ line can be used to estimate the BH mass as a strong correlation between \ha\ and \hb\ linewidths has been found in low-redshift quasars \citep{shen_catalog_2011,greene_estimating_2005}. We therefore also use the following scaling relation \citep{shen_catalog_2011},
\begin{equation}\label{eq:ha}
    \begin{split}
    \log \left( \frac{M_{\rm BH, H\alpha}}{M_{\odot}} \right)
    = 0.379 & + 0.43 \log \left( \frac{L_{\rm H\alpha}}{10^{42}\ {\rm erg\ s}^{-1}} \right) \\
    & + 2.1 \log \left( \frac{{\rm FWHM}_{\rm H\alpha}}{{\rm km\ s}^{-1}} \right),
    \end{split}
\end{equation}
to estimate the BH mass from the FWHM and the line luminosity of the \ha\ line. This relation has a systematic scatter of $\sim 0.3\ {\rm dex}$ \citep{shen_catalog_2011} (up to $\sim 0.6\ {\rm dex}$ if reverberation mapping uncertainties are taken into account).

\subsubsection{\ion{Fe}{2}-corrected \hb}

Four out of the five quasars in our sample exhibit strong iron emission in their rest-optical spectra. In light of the Eigenvector I relation found in low-redshift quasars \citep{sulentic_eigenvector_2000}, which indicates an anticorrelation between the FWHM of the \hb\ line and the strength of the \ion{Fe}{2} emission in optical quasar spectra, \cite{du_radiusluminosity_2019} established a scaling relation for BLR size,
\begin{equation}\label{eq:fehb1}
    \begin{split}
        \log \left( \frac{R_{\rm BLR, H\beta}}{\rm ltd} \right)
        = 1.65 & + 0.45 \log 
        \left( \frac{L_{5100}}{10^{44} {\rm erg\ s}^{-1}}\right) \\
        & - 0.35\, R_{\mathrm{Fe,H}\beta},
    \end{split}
\end{equation}
which includes the iron-correction factor $R_{\mathrm{Fe,H}\beta}=F_{\rm Fe\ {\small II}}/F_{\rm H\beta}$, defined as the ratio of the integrated \ion{Fe}{2} flux between $4434\ {\rm \AA}$ and $4684\ {\rm \AA}$, and the broad \hb\ line flux. Note that this relation has a systematic scatter of $\sim 0.5\ {\rm dex}$. Subsequently, we use the virial relation to obtain the BH mass from this estimator,
\begin{equation}\label{eq:fehb2}
    M_{\rm BH} = f\frac{R_{\rm BLR, H\beta} {\rm FWHM}_{\rm H\beta}^2 }{G},
\end{equation}
where $G$ is the gravitational constant and $f$ is the virial factor related to the geometry and orientation of the BLR. In this study, our goal is to explore how the estimated BH mass from \hb\ is affected by the variance in the iron emission in our sample, which introduces a correction to the assumed relation between the radius of the broad line region and the quasar luminosity. Therefore, we calibrate $f$ for each quasar such that the estimated BH masses from \cref{eq:hb} and \cref{eq:fehb1,eq:fehb2} agree for $R_{\mathrm{Fe,H}\beta}=0$ (i.e for the canonical $R_{\rm BLR}-L$ relation).

\subsection{Measured black hole masses}

We summarize our BH mass measurements, quoting the median and the $16$th and $84$th percentiles of the MCMC posterior distribution, in \cref{tab:MBH}, including the measured iron-correction factor $R_{\mathrm{Fe,H}\beta}$ and the calibration for the virial factor $f$ used in \cref{eq:fehb2}, alongside the most relevant measurements from the literature. We choose to only compare our results to \mgii-based BH mass measurements as far as the rest-UV spectrum is concerned, as measurements based on the \civ\ emission line are less reliable due to the extreme \civ\ properties seen in high-redshift quasars \citep{schindler_x-shooteralma_2020}. Also note that SDSS J0100+2802 is the only quasar in our sample that has an existing \hb\ measurement \citep{eilers_eiger_2023}.

\begin{table*}[t!]
\centering
\caption{New black hole mass measurements from the rest-optical spectra of the quasars in our sample, in comparison to existing $M_{\rm BH}$ measurements from the literature: E23 - \cite{eilers_eiger_2023}; E20 - \cite{eilers_detecting_2020}; E18 - \cite{eilers_first_2018}. We also include the measured iron-correction factor $R_{\mathrm{Fe,H}\beta}$ as well as the calibrated value for the virial factor $f$ in \cref{eq:fehb2} that was used to calculate the \ion{Fe}{2}-corrected masses. The quoted uncertainties reflect the $16$th and $84$th percentiles of the posterior distribution from the MCMC, and so do not include the aforementioned systematic uncertainties.}\label{tab:MBH}
\begin{tabular}{lccccc|ccc@{\hskip 5pt}c}
\hline\hline
Quasar 
& \multicolumn{5}{c|}{new $\log_{10}M_{\rm BH}/$\msol}
& \multicolumn{3}{c}{$\log_{10}M_{\rm BH}/$\msol\ from lit.} \\
& H$\alpha$ 
& H$\beta$ 
& H$\beta$ (Fe\,\textsc{ii}-corr.) 
& $R_{\mathrm{Fe,H}\beta}$
& $f$ 
& Mg\,\textsc{ii} 
& H$\beta$ 
& ref. \\
\hline
SDSS J0100+2802 & $9.82_{-0.01}^{+0.01}$ & $10.28_{-0.01}^{+0.01}$ & $10.11_{-0.01}^{+0.01}$ & $0.48_{-0.01}^{+0.01}$ & $1.34$ & $9.7\pm0.1$ & $10.2\pm0.1$ & E23\\
PSO J158--14 & $9.31_{-0.01}^{+0.01}$ & $9.54_{-0.01}^{+0.01}$ &  $9.26_{-0.01}^{+0.01}$ & $0.81_{-0.02}^{+0.01}$ & $1.24$ & $9.20\pm0.03$ & -- & E20\\
SDSS J1335+3533 & $9.16_{-0.01}^{+0.01}$ & $9.36_{-0.03}^{+0.03}$ & $9.03_{-0.03}^{+0.03}$ & $0.95_{-0.01}^{+0.02}$ & $1.20$ & $9.62_{-0.57}^{+0.55}$ & -- & E18\\
CFHQS J2100--1715 & $8.54_{-0.03}^{+0.03}$ & $8.78_{-0.03}^{+0.02}$ & $8.43_{-0.05}^{+0.05}$ & $1.05_{-0.08}^{+0.08}$ & $1.15$ & $9.34\pm0.04$ & -- & E20\\
CFHQS J2229+1457 & $8.86_{-0.01}^{+0.01}$ & $8.91_{-0.02}^{+0.02}$ & $8.92_{-0.02}^{+0.02}$ & $0.07_{-0.02}^{+0.02}$ & $1.12$ & $9.16_{-0.08}^{+0.07}$ & -- & E20\\
\hline
\end{tabular}

\end{table*}

\begin{table*}[t!]
\centering
\caption{Measurements of the continuum luminosity at rest-frame $5100\ {\rm \AA}$ as well as Eddington ratios, $\lambda_{\rm Edd}$, corresponding to the three BH mass measurements from \cref{tab:MBH}. The assumed bolometric correction is $9.26$ \citep{Richards_2006}. The quoted uncertainties reflect the $16$th and $84$th percentiles of the posterior distribution from the MCMC, and so do not include the systematic uncertainties of the BH mass measurements.}\label{tab:edd}
\begin{tabular}{lcccc@{\hskip 5pt}c}
\hline\hline
Quasar 
& $\log_{10}L_{5100}$ 
& $\lambda_{\rm Edd}^{\rm H\alpha}$
& $\lambda_{\rm Edd}^{\rm H\beta}$ 
& $\lambda_{\rm Edd}^{\rm H\beta, Fe\,\textsc{ii}}$ \\
& [erg s$^{-1}$] & & & \\
\hline
SDSS J0100+2802 & $47.15_{-0.01}^{+0.01}$ & $1.56_{-0.03}^{+0.01}$ & $0.54_{-0.01}^{+0.02}$ & $0.80_{-0.02}^{+0.02}$ \\
PSO J158--14 & $46.42_{-0.01}^{+0.01}$ & $0.94_{-0.01}^{+0.01}$ &  $0.55_{-0.01}^{+0.01}$ & $1.05_{-0.01}^{+0.01}$\\
SDSS J1335+3533 & $46.12_{-0.01}^{+0.01}$ & $0.66_{-0.01}^{+0.01}$ & $0.42_{-0.02}^{+0.03}$ & $0.90_{-0.05}^{+0.07}$\\
CFHQS J2100--1715 & $45.11_{-0.01}^{+0.01}$ & $0.27_{-0.01}^{+0.02}$ & $0.16_{-0.01}^{+0.01}$ & $0.35_{-0.04}^{+0.05}$\\
CFHQS J2229+1457 & $44.87_{-0.01}^{+0.01}$ & $0.08_{-0.01}^{+0.01}$ & $0.07_{-0.01}^{+0.01}$ & $0.07_{-0.01}^{+0.01}$\\
\hline
\end{tabular}
\end{table*}

All of our new BH mass measurements are consistent with one another and previously published BH mass measurements from the literature within systematic uncertainties (up to $\sim 0.6\ {\rm dex}$). This includes a spot-on agreement of our \hb\ mass with the one previously measured by \cite{eilers_eiger_2023} for SDSS J0100+2802 using NIRCam Wide Field Slitless Spectroscopy observations. In all cases, \hb-derived BH masses are systematically higher than masses derived from \ha. Notably, the iron-correction decreases the derived BH masses with respect to the canonical \hb\ measurements by up to $\sim0.4$ dex, producing masses more consistent with (and in some cases even lower than) the \ha-derived measurements. Our derived iron-correction factors are significant but still lower than in some of the quasars presented by \cite{yang_spectroscopic_2023} -- taking the iron correction into account would therefore result in an even more significant decrease to the \hb-based BH masses for those objects.

In \cref{tab:edd}, we also include the measurement of the continuum luminosity at rest-frame $5100\ {\rm \AA}$, $L_{5100}$, which we use to derive the Eddington ratios provided in the same table. We compute each Eddington ratio as a ratio of the bolometric luminosity and the Eddington luminosity corresponding to our estimated BH masses from all three methods, $\lambda_{\rm Edd} = L_{\rm bol}/L_{\rm Edd}$. We employ the bolometric correction of $9.26$ \citep{Richards_2006}. As can be seen, the Eddington ratios derived from the three different BH mass estimates vary significantly (even though the BH mass measurements are consistent with one another within the systematic uncertainties), and so one should use caution when interpreting them.

Overall, our new measurements confirm the existence of billion-solar-mass BHs at the centers of high-redshift quasars with small proximity zones, pressing on the question of their fast assembly in the early Universe.

\section{Quasar lifetimes from extended emission}\label{sec:nebulae}

To investigate the problem of growing billion-solar-mass BHs in the short lifetimes implied by their line-of-sight proximity zones \citep{eilers_first_2018,davies_constraining_2020,eilers_detecting_2021} as well as their \lya\ nebulae \citep{durovcikova_quasar_2025}, we use the deep NIRSpec IFU observations presented here to obtain new lifetime measurements using their rest-frame optical extended emission. We implement similar methods as by \citep{durovcikova_quasar_2025}. For all quasars in this sample, we focus on searching for extended emission around \ha\ and \oiii$\lambda 5007$ as these nebulae are expected to be the brightest in the rest-optical wavelengths and thus the most likely cases where the edge of the nebula can possibly be detected. In this work, we focus on the spatial extent of the nebulae. The analysis of the kinematics revealed by the extended emission in these quasars will be presented by Ishikawa et al. (2025, in preparation).

\subsection{PSF subtraction}\label{sec:PSF}

In order to search for the faint, extended emission around the quasars in our sample, we first need to subtract the point-like emission from the nuclear region of the quasar. We extract the point-spread function (PSF) directly from the data as follows. As the JWST PSF varies significantly across wavelength, we perform separate PSF subtraction for each emission line nebula of interest. In each case, we construct the PSF model by collapsing the data cube across wavelength channels corresponding to the wings of the broad emission line components, as the high velocity broad-line emission is thought to arise in the nuclear region and thus resemble a true point source. To minimize systematics due to wavelength changes of the PSF, we do this both for the blue and the red wings of the emission line and then average them to compute the resultant PSF model. We display the corresponding wavelength regions in Appendix \ref{app:specregions} for each quasar and each emission line of interest. During this PSF construction, we take care to exclude spectral channels that are contaminated by residual MSA leaks or artifacts that we were unable to remove during the data reduction and preprocessing stage. In the case of both \ha\ and \oiii, we checked that our results are robust against the exact wavelength windows chosen to construct the PSF.

We subtract the PSF model from each spectral channel around the given emission line, normalizing it separately in each channel to the aperture sum with a radius of two pixels (corresponding to $0.2''$) centered at the quasar location. Again, since residual MSA leaks and other artifacts are a concern for our science goal, we mask the largest outliers in each PSF-subtracted channel before proceeding to the next step.

\subsection{Search for extended emission}\label{sec:search}

In order to identify nebular emission, we use the PSF-subtracted cube to search for a group of connected voxels (volume pixels) above a certain signal-to-noise (SNR) threshold in the vicinity of the quasar. We first compute a smoothed cube, following \cite{hennawi_quasars_2013,arrigoni_battaia_deep_2015,farina_mapping_2017,farina_requiem_2019,durovcikova_quasar_2025},
\begin{equation}
    {\rm SMOOTH}[\chi_{x,y,\lambda}] = \frac{{\rm CONVOL}[{\rm DATA}_{x,y,\lambda}-{\rm PSF}_{x,y,\lambda}]}{\sqrt{{\rm CONVOL}^2[\sigma^2_{x,y,\lambda}]}},
\end{equation}
where ${\rm DATA}_{x,y,\lambda}$ represents the data cube, ${\rm PSF}_{x,y,\lambda}$ is the aforementioned 3-dimensional PSF model normalized at each spectral channel, and $\sigma^2_{x,y,\lambda}$ represents the variance data cube. The ${\rm CONVOL}$ operation denotes a convolution with a 3-dimensional Gaussian kernel. We choose the standard deviation of the kernel to be $\sigma_{x,y}=0.1''$ in the spatial direction (corresponding to one spatial pixel), however, since this dataset has a lower spectral resolution than the MUSE dataset analyzed by \cite{durovcikova_quasar_2025}, we choose to not apply any smoothing in the spectral direction. The thus constructed ${\rm SMOOTH}[\chi_{x,y,\lambda}]$ is essentially a smoothed version of a data cube representing the SNR of each voxel. 

\begin{table*}[t!]
\centering
\caption{A summary of \ha\ and \oiii\ nebular (non-)detections and their properties: the $5\sigma$ detection limit over a $1{\rm arcsec}^2$ aperture, ${\rm SB}^1_{5\sigma}$, the maximum extent of the nebula, $d_{\rm max}^{\rm Neb}$, and the corresponding lifetime measurement, $\log{t_{\rm Q}^{\rm Neb}}$.}\label{tab:lifetimes}
\begin{tabular}{lccc ccc@{\hskip 5pt}c}
\hline\hline
Quasar & ${\rm SB}^1_{5\sigma,{\rm H\alpha}}$
& $d_{\rm max}^{\rm H\alpha}$
& $\log{t_{\rm Q}^{\rm H\alpha}}$
& ${\rm SB}^1_{5\sigma,{\rm \oiii}}$
& $d_{\rm max}^{\rm \oiii}$
& $\log{t_{\rm Q}^{\rm \oiii}}$ \\
& $[{\rm erg\ s^{-1}\ cm^{-2}\ arcsec^{-2}}]$ & $[{\rm pkpc}]$ & $[{\rm yr}]$ & $[{\rm erg\ s^{-1}\ cm^{-2}\ arcsec^{-2}}]$ & $[{\rm pkpc}]$ & $[{\rm yr}]$ \\
\hline
SDSS J0100+2802 & $3.0\times 10^{-18}$ & $8.96_{-1.79}^{>1.79}$ & $4.47_{-0.10}^{>0.08}$ & $2.5\times 10^{-18}$ & $7.77_{-2.99}^{>2.99}$ & $4.40_{-0.21}^{>0.14}$ \\
PSO J158--14 & $3.2\times 10^{-18}$ & $>11.01$ &  $>4.56$ & $2.6\times 10^{-18}$ & $2.36_{-0.91}^{+2.34}$ & $3.89_{-0.21}^{>0.30}$ \\
SDSS J1335+3533 & $2.2\times 10^{-18}$ & $4.14_{-1.80}^{+0.83}$ & $4.13_{-0.25}^{+0.08}$ & $1.9\times 10^{-18}$ & $1.84_{-0.37}^{+0.37}$ & $3.77_{-0.10}^{+0.08}$ \\
CFHQS J2100--1715 & $1.8\times 10^{-18}$ & $<1.45$ & $<3.67$ & $1.4\times 10^{-18}$ & $<1.45$ & $<3.67$ \\
CFHQS J2229+1457 & $2.3\times 10^{-18}$ & $4.86_{-0.00}^{+2.14}$ & $4.20_{-0.00}^{+0.16}$ & $1.9\times 10^{-18}$ & $7.28_{-0.00}^{+0.00}$ & $4.38_{-0.00}^{+0.00}$ \\
\hline
\end{tabular}
\end{table*}

With this smoothed data cube at hand, we first search for the voxel with the highest SNR in a 1000\,{\rm km/s} spectral window around the emission line of interest and at most $0.5''$ away from the quasar. We then run a friends-of-friends algorithm to connect to it all voxels that are above a certain SNR threshold, ${\rm SNR_{\rm thres}}$, and lie within a linking distance both spatially, $l_{{\rm thres},x,y}$, and spectrally, $l_{{\rm thres},\lambda}$. We also require that at least $30$ voxels need to be connected over at least two spectral channels to establish the detection of a nebula. Due to the clumpiness of some of the nebular detections, particularly in the \oiii\ transition, we run the friends-of-friends algorithm around multiple significant pixels to link up parts of the extended emission which are spatially more dispersed. We repeat this linking procedure for a range of parameters: ${\rm SNR_{\rm thres}} = \{ 2.0, 3.0 \}$, $l_{{\rm thres},x,y} = \{ 0.2'', 0.3'' \}$, and $l_{{\rm thres},\lambda} = \{ 18 {\rm \AA}, 36 {\rm \AA} \}$ (corresponding to a linking length of 2 and 3 spatial pixels and 1 and 2 spectral pixels of the medium-resolution grating, respectively); then we use this collection of 3D nebular masks to compute a median mask of the nebula for each quasar and each emission line of interest. Note that we keep these linking parameters consistent across our sample and apply the same spectral linking length to the high-resolution grating observations of SDSS J0100+2802.

We compile the results of our nebular search in \cref{fig:nebJ0100,fig:nebJ158,fig:nebJ1335,fig:nebJ2100,fig:nebJ2229} for both \ha\ (top row, orange color scheme) and \oiii\ nebulae (bottom row, green color scheme). In all cases, we show pseudo-narrowband images corresponding to the quasar emission (panel 1), the rescaled PSF model (panel 2), the PSF-subtracted emission (panel 3), and the $\chi$ (panel 4) and smoothed $\chi$ (panel 5) image collapsed over the wavelength range corresponding to the median nebular mask, or over a $1000\,{\rm km/s}$ spectral window in the case of non-detections. In panel 5, we also show the contours corresponding to the median nebular mask.

\subsection{Measuring quasar lifetimes}

The measurement of quasar lifetimes from their extended emission is based on the size of the region that the quasar has been able to photoionize during its active phase. Hence, our aim is to measure the maximum transverse (projected) distance between the quasar and the edge of the nebula, and relate it to the quasar lifetime via the speed of light, $c$,
\begin{equation}
    t_{\rm Q}^{\rm Neb} = \frac{d_{\rm max}^{\rm Neb}}{c}.
\end{equation}
We use the projected size of the nebula instead of the three-dimensional distance as the position and velocity information are difficult to disentangle along our line of sight (due to peculiar velocities) and we do not have information about the ionization cone opening angle. This method also assumes that once the quasar begins its activity, the size of the photoionized region grows at or close to the speed of light. In the case of \ha, the same arguments as for \lya\ are valid -- the lightspeed travel of ionizing photons through the CGM is enabled by the highly ionized nature of the CGM together with the typically very low volume filling factor of the cool, neutral gas clouds responsible for this nebular emission \citep{prochaska_quasars_2009,mccourt_characteristic_2018,pezzulli_high_2019}, which make ionization-front effects on the timescales negligible \citep{durovcikova_quasar_2025}. The forbidden \oiii$\lambda 5007$ transition traces rarified \ion{O}{1} or \ion{O}{2} gas that has been photoionized by the quasar, and also traces the cooler and clumpier phase of the CGM \citep[e.g.][]{liu_observations_2013}. Using the speed of light to derive the quasar lifetime also assumes that the time delay between photoionization and photon emission is negligible, which is justified in Appendix \ref{app:delaytime}.

\begin{figure*}
    \centering
    \includegraphics[width=\linewidth,trim={0 0.2cm 0 0.2cm},clip]{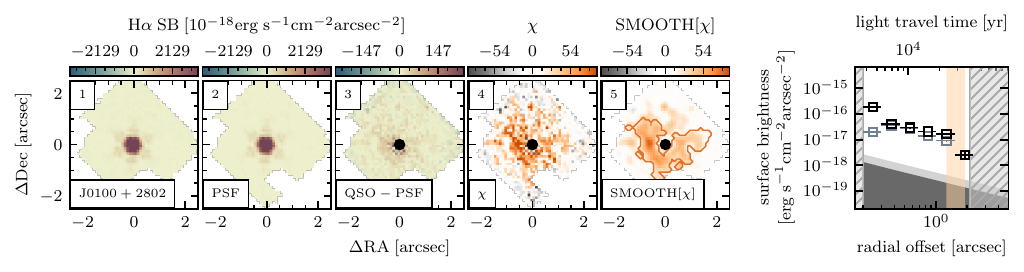}
    \includegraphics[width=\linewidth,trim={0 0.2cm 0 0.2cm},clip]{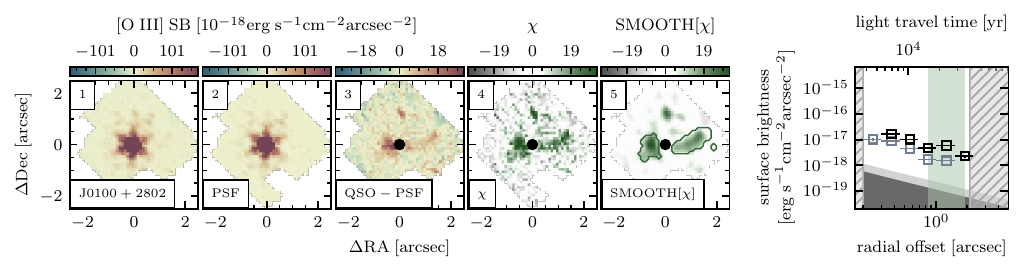}
    \caption{The results of our \ha\ (top row, orange color scheme) and \oiii\ (bottom row, green color scheme) nebular search for J0100+2802. Left column: From left to right, the panels display pseudo-narrowband images of 1) the data cube collapsed around the emission line in units of surface brightness, 2) the PSF extracted from the broad wings of the emission line, 3) the PSF-subtracted equivalent of panel 1, 4) the $\chi$ data, essentially representing the SNR of the PSF-subtracted data, and 5) the smoothed $\chi$ data showing the extended nebular emission. Panels 1, 3, 4, and 5 are collapsed across the wavelength range of the detected nebula, and panel 2 is collapsed across the wavelengths of the PSF extraction as displayed in Appendix \ref{app:specregions}. White patches correspond to masked foregrounds/artifacts, and the black circle marks the position of the quasar. Right column: The annulus-averaged surface brightness profiles extracted from panel 3, computed both using the full pseudo-narrowband image (gray data points) and only taking into account the nebular voxels (black data points). The black data points are used to identify the maximum extent of the nebula, from which the lifetimes are derived. The noise floor (gray shaded regions corresponding to $1\sigma$ and $2\sigma$) is extracted from the outermost annulus, excluding pixels identified as part of the nebula. Additionally, the hatched vertical regions mark the central region inaccessible due to PSF residuals as well as the region that lies beyond the field of view of the NIRSpec IFU.}
    \vspace{0.5cm}
    \label{fig:nebJ0100}
\end{figure*}

\begin{figure*}
    \centering
    \includegraphics[width=\linewidth,trim={0 0.2cm 0 0.2cm},clip]{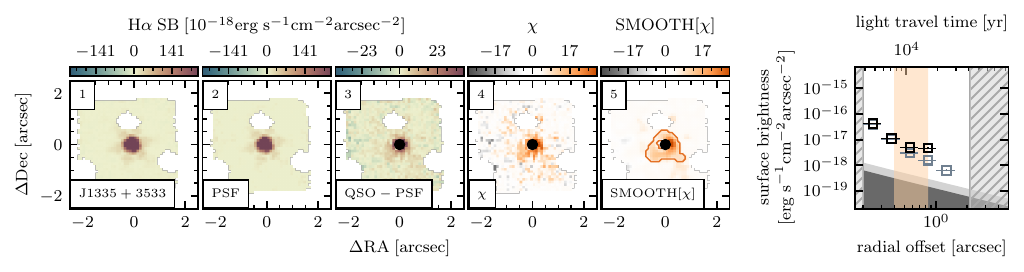}
    \includegraphics[width=\linewidth,trim={0 0.2cm 0 0.2cm},clip]{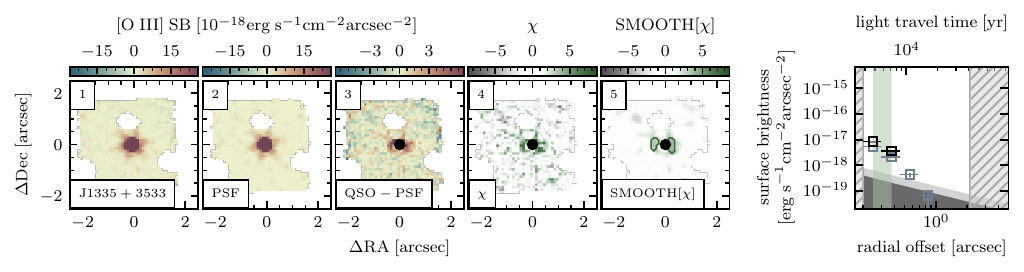}
    \caption{Same as \cref{fig:nebJ0100}, but for J1335+3533.}
    \label{fig:nebJ1335}
\end{figure*}

\begin{figure*}
    \centering
    \includegraphics[width=\linewidth,trim={0 0.2cm 0 0.2cm},clip]{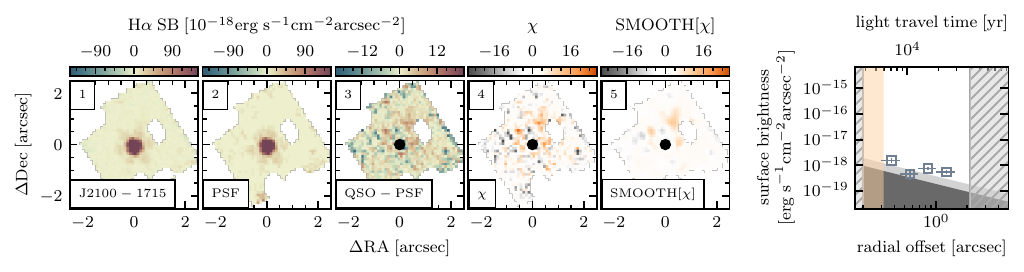}
    \includegraphics[width=\linewidth,trim={0 0.2cm 0 0.2cm},clip]{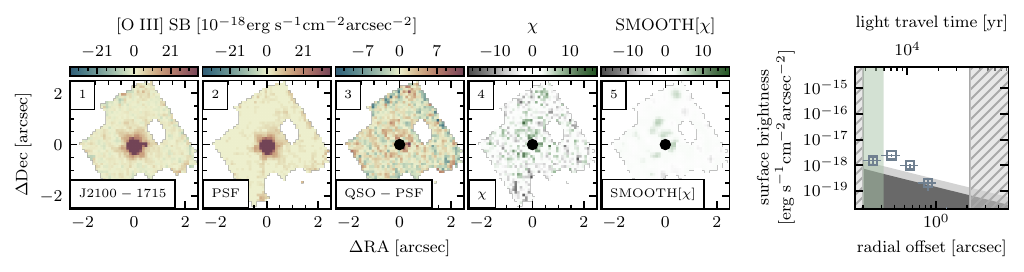}
    \caption{Same as \cref{fig:nebJ0100}, but for J2100--1715. Note that since these are both non-detections, panels 1, 3, 4, and 5 are now collapsed in a $1000\,{\rm km/s}$ spectral window around the emission line.}
    \label{fig:nebJ2100}
\end{figure*}

To measure the maximum projected size of each detected nebula, we compute its corresponding surface brightness profile (shown in the right column of \cref{fig:nebJ0100,fig:nebJ158,fig:nebJ1335,fig:nebJ2100,fig:nebJ2229}). To do this, we collapse the PSF-subtracted data cube across the wavelength range of the detected nebular emission and perform aperture sum in logarithmically-spaced annular regions centered on the quasar at increasing radii, starting at a separation of $0.2''$ away from the quasar to avoid PSF residuals. Following \cite{durovcikova_quasar_2025}, we compute the surface brightness profile in two ways: a) by only taking into account the voxels identified by the nebular mask (to avoid diluting signal at the edges of the nebula, black data points in \cref{fig:nebJ0100,fig:nebJ158,fig:nebJ1335,fig:nebJ2100,fig:nebJ2229}), and b) by computing the aperture sum within the whole annular region irrespective of the nebular mask \citep[e.g. following][shown as gray data points]{borisova_ubiquitous_2016,farina_requiem_2019,arrigonibattaia_qso_2019}. 
We also display the background surface brightness noise in gray, which we compute as the variance of the surface brightness of pixels in the outermost annulus that are not included in the nebular mask. Note that the displayed surface brightness profiles are not corrected for cosmological dimming. The maximum extent of the nebula, $d_{\rm max}^{\rm Neb}$, is identified based on the largest radius bin with a nebular detection (using method a), i.e. black data points), and this procedure is repeated for all nebular masks identified using our parameter search described in \S~\ref{sec:search} to obtain a distribution of nebular sizes and corresponding lifetimes.

In addition to measuring the maximum extent of the nebulae, we also quantify the corresponding surface brightness limits by collapsing each data cube across five wavelength channels centered at the quasar's emission line of interest and measuring the variance in a $1{\rm arcsec}^2$ aperture, following the literature \citep[e.g.][]{farina_mapping_2017,farina_requiem_2019,durovcikova_quasar_2025}. The measured $5\sigma$ limits are also given in \cref{tab:lifetimes}.

In what follows, we present and discuss the nebular (non-)detections for each quasar in our sample separately, and summarize the results in \cref{tab:lifetimes}.

\subsubsection{SDSS J0100+2802}\label{sec:J0100}

The quasar J0100+2802 is known as the most luminous quasar hosting the most massive SMBH at $z>6$. This quasar is included in the sample of \cite{durovcikova_quasar_2025}, and its \lya\ nebular lifetime was constrained to be $\log{t_{\rm Q}^{\rm Ly\alpha}/{\rm yr}} = 4.77_{-0.10}^{+0.08}$, in agreement with its line-of-sight proximity-zone lifetime, $\log{t_{\rm Q}^{R_{\rm p}}/{\rm yr}} = 5.1^{+1.3}_{-0.7}$ \citep[\cref{tab:sample};][]{davies_constraining_2020}, which extends beyond the field of view of the NIRSpec IFU. Very recently, this quasar's transverse proximity effect (its ``light echo'') has also been detected for the first time and yielded a lifetime measurement of $\log{t_{\rm Q}/{\rm yr}} = 5.6_{-0.3}^{+0.1}$ \citep{eilers_light_2025}.

We clearly detect nebular emission in both \ha\ and \oiii\ (\cref{fig:nebJ0100}), yielding $\log{t_{\rm Q}^{\rm H\alpha}/{\rm yr}} = 4.47_{-0.10}^{>0.08}$ and $\log{t_{\rm Q}^{\rm \oiii}/{\rm yr}} = 4.40_{-0.21}^{>0.14}$. Note that the upper bounds on our measurements are lower limits. \ha\ emission is extended and likely continues beyond the field of view of these observations, in agreement with the lifetime constraints from \lya. \oiii\ emission is localized around what could be an axisymmetrical region and shows an interesting velocity structure and a tentative companion galaxy (Ishikawa et al. 2025, in preparation). Overall, this quasar is thus a unique example for which four independent lifetime measurements consistently point towards an accretion timescale of less than a million years.

\subsubsection{PSO J158--14}\label{sec:J158}

The small line-of-sight proximity zone of this quasar 
\citep[\cref{tab:sample};][]{eilers_detecting_2021} has recently been discovered to be truncated by an extremely metal-poor absorption system \citep{durovcikova_extremely_2025}. This finding has been corroborated by the detection of a large \lya\ nebula, yielding a nebular lifetime of at least $\log{t_{\rm Q}^{\rm Ly\alpha}/{\rm yr}} = 4.89_{-0.21}^{+0.30}$ \citep{durovcikova_quasar_2025}, much larger than it would be expected from its observed proximity zone. Despite its ``older'' nature, we still include this quasar in our sample to check for consistency with the previous nebular lifetime measurement from \lya.

\begin{figure*}
    \centering
    \includegraphics[width=\linewidth,trim={0 0.2cm 0 0.2cm},clip]{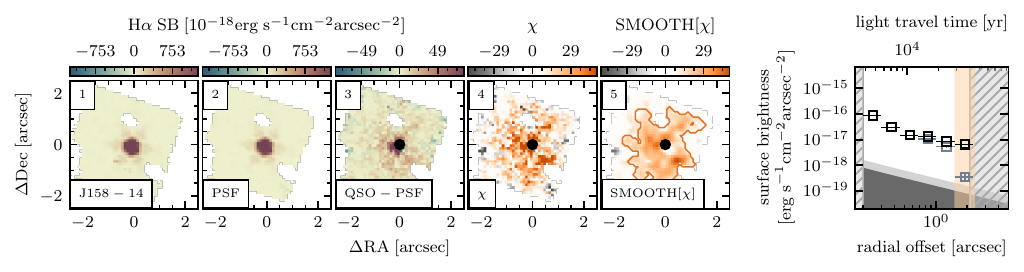}
    \includegraphics[width=\linewidth,trim={0 0.2cm 0 0.2cm},clip]{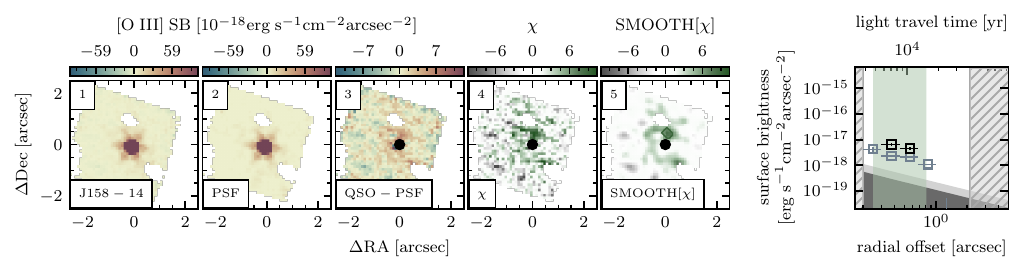}
    \caption{Same as \cref{fig:nebJ0100}, but for J158--14.}
    \label{fig:nebJ158}
\end{figure*}

The corresponding \ha\ extended emission is clearly detected and occupies almost the full field of view of the NIRSpec IFU (\cref{fig:nebJ158}), yielding $\log{t_{\rm Q}^{\rm H\alpha}/{\rm yr}} >4.56$, as expected based on its \lya\ nebula. This is the case for all parameter combinations explored in our nebular search, unabiguously confirming that this quasar has been active for much longer than its line-of-sight proximity zone seems to suggest. On the other hand, the \oiii\ extended emission is clearly present, however, it is difficult to determine its full extent due to its clumpiness and lower SNR, thus resulting in a shorter lifetime constraint which we interpret as a lower limit, $\log{t_{\rm Q}^{\rm \oiii}/{\rm yr}} = 3.89_{-0.21}^{>0.30}$. We also note that this system shows rotating kinematics, with a likely host galaxy detection, which is presented in detail in Ishikawa et al. 2025 (in preparation).

\subsubsection{SDSS J1335+3533}\label{sec:J1335}

J1335+3533 is a weak-line quasar, whose line-of-sight proximity zone yields a lifetime constraint of only $\log{t_{\rm Q}^{R_{\rm p}}/{\rm yr}} = 3.0 \pm 0.4$ \citep[\cref{tab:sample};][]{eilers_first_2018}. This quasar does not have existing deep rest-UV IFU observations and hence the properties of its \lya\ nebula are currently unknown.

We detect both an \ha\ and a fainter \oiii\ nebulae in this quasar, both very compact, spanning $< 1''$ in their spatial extent and exhibiting similar morphologies. Both of these detections are well above the sensitivity of these observations, clearly displaying the edges of the nebulae and yielding lifetime constraints that are slightly larger than its proximity-zone lifetime, namely $\log{t_{\rm Q}^{\rm H\alpha}/{\rm yr}} = 4.13_{-0.25}^{+0.08}$ and $\log{t_{\rm Q}^{\rm \oiii}/{\rm yr}} = 3.77_{-0.10}^{+0.08}$. Nevertheless, the extended emission still suggests that this quasar has been active for $\lesssim 10^4\ {\rm yr}$, confirming its extremely young nature.

\begin{figure*}
    \centering
    \includegraphics[width=\linewidth,trim={0 0.2cm 0 0.2cm},clip]{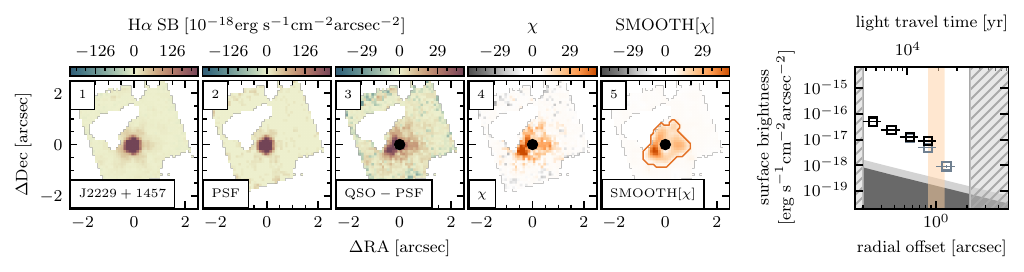}
    \includegraphics[width=\linewidth,trim={0 0.2cm 0 0.2cm},clip]{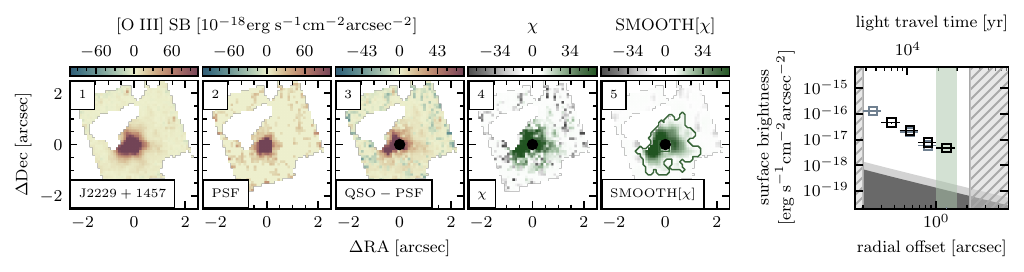}
    \caption{Same as \cref{fig:nebJ0100}, but for J2229+1457.}
    \label{fig:nebJ2229}
\end{figure*}

\subsubsection{CFHQS J2100--1715}\label{sec:J2100}

J2100--1715 was the first small-proximity-zone quasar \citep[$\log{t_{\rm Q}^{R_{\rm p}}/{\rm yr}} = 2.3 \pm 0.7$, \cref{tab:sample};][]{eilers_detecting_2021} to be searched for an extended \lya\ emission using a deep MUSE observation, resulting in a non-detection \citep{farina_requiem_2019,drake_ly_2019}. Its MUSE observations were reanalyzed by \cite{durovcikova_quasar_2025}, confirming this non-detection and deriving an upper limit on its transverse quasar lifetime to be $\log{t_{\rm Q}^{\rm Ly\alpha}/{\rm yr}}< 3.88$, in agreement with its line-of-sight lifetime.

Our search for \ha\ and \oiii\ nebulae in deep NIRSpec IFU observations resulted in a non-detection in both cases, tightening the upper limit on this quasar's nebular lifetime to $\log{t_{\rm Q}^{\rm H\alpha}/{\rm yr}} = \log{t_{\rm Q}^{\rm \oiii}/{\rm yr}} < 3.67$. Intriguingly, this quasar also has the highest iron-correction factor (see \cref{tab:MBH}) among our sample, which has been shown to correlate with the Eddington ratio in Type~I quasars \citep{shen_diversity_2014}. Our derived BH mass and luminosity do not suggest a particularly large Eddington ratio (\cref{tab:edd}), however, it should be emphasized that the validity of a uniform bolometric correction used to derive these is questionable, particularly in the case of atypical quasars like this one. Its high iron-correction factor together with the extremely short nebular and proximity-zone lifetimes could indeed suggest the presence of high accretion rates in this quasar. The agreement in the transverse (both rest-optical and rest-UV) and line-of-sight lifetimes also suggests that (past) sightline-dependent obscuration effects are unlikely the cause of this extremely small proximity zone, supporting the finding of \cite{durovcikova_quasar_2025}.

\subsubsection{CFHQS J2229+1457}\label{sec:J2229}

The proximity zone of J2229+1457 is one of the smallest proximity zones measured at high redshift, implying a line-of-sight lifetime of only $\log{t_{\rm Q}^{R_{\rm p}}/{\rm yr}} = 2.9^{+0.8}_{-0.9}$ \citep[\cref{tab:sample};][]{eilers_detecting_2021}. This quasar was also included in the sample of \cite{durovcikova_quasar_2025} and resulted in a non-detection of a \lya\ nebula and a corresponding nebular lifetime limit of $\log{t_{\rm Q}^{\rm Ly\alpha}/{\rm yr}}< 3.88$, in agreement with its proximity-zone lifetime.

Our nebular search reveals booming nebulae in both \ha\ and \oiii\ (\cref{fig:nebJ2229}), both with very similar morphologies and extents, yielding $\log{t_{\rm Q}^{\rm H\alpha}/{\rm yr}} = 4.20_{-0.00}^{+0.16}$ and $\log{t_{\rm Q}^{\rm \oiii}/{\rm yr}} = 4.38_{-0.00}^{+0.00}$, at about $\sim2\sigma$ different than the lifetime implied by its proximity zone and its \lya\ nebular non-detection. J2229+1457 is the only quasar in this sample whose \oiii\ emission, both its nuclear (narrow-line) and its extended emission, is brighter than its \ha\ line and nebula, suggesting that the ionization conditions in this system are qualitatively different from the rest of the sample (also noticable in the very weak nuclear iron emission compared to the other quasars, see \cref{fig:spectral_fits}, as well as its extremely low Eddington ratio, see \cref{tab:edd}). The edge of the nebula is detected with very high significance, particularly in \oiii, where even changing our nebular search parameters does not affect its measured size.

\begin{figure*}
    \centering
    \includegraphics[width=\linewidth]{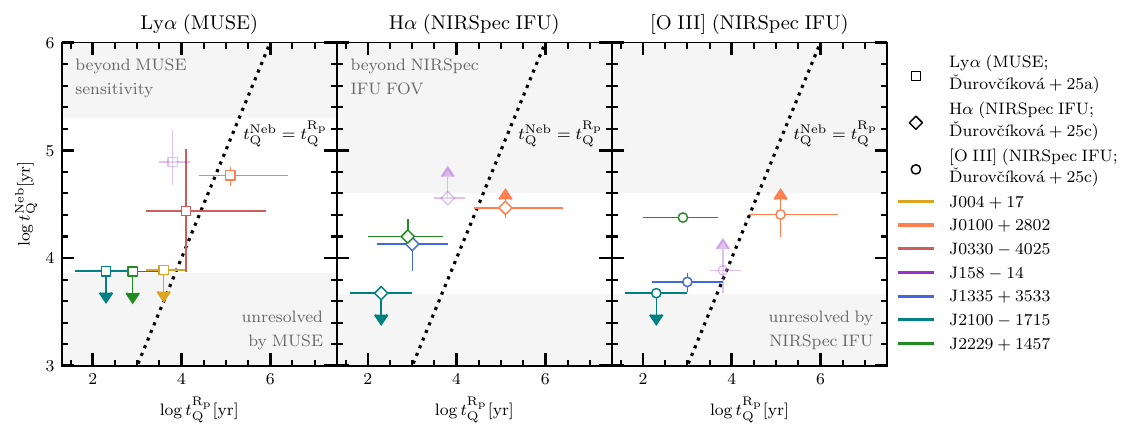}
    \caption{A summary of nebular and proximity-zone lifetime constraints from this work (NIRSpec IFU, the middle panel displays \ha\ measurements, and the right panel displays \oiii\ measurements) and the literature \citep[MUSE, left panel;][]{durovcikova_quasar_2025}. The proximity zone measurements are taken from \cite{eilers_first_2018,davies_constraining_2020,eilers_detecting_2021}, and in all three panels, we include the dashed 1:1 line to guide the eye. The gray shaded regions show regions that are either beyond the spatial resolution of the corresponding instrument, or regions that are beyond detection due to the sensitivity or the field of view (FOV) of the instrument. The faded, purple set of data points correspond to J158--14 whose sightline has been found to be contaminated by a metal-poor absorption system \citep{durovcikova_extremely_2025}, thereby truncating its proximity zone and causing a large discrepancy between its line-of-sight and transverse (nebular) lifetime measurements. All other data points confirm the existence of quasars with extremely short UV-luminous lifetimes at high redshifts.}
    \label{fig:tqsummary}
\end{figure*}

Note the presence of a foreground galaxy at $\sim 1''$ to the North-East of the quasar (masked in \cref{fig:nebJ2229}, but for more details see Ishikawa et al. 2025, in preparation). \cite{yue_detecting_2023} showed that this quasar is unlikely to be lensed, based on its PSF in Hubble Space Telescope imaging data, however, a magnification of $<2$ could still be present based on the Singular Isothermal Sphere (SIS) lensing model. Even if this magnification were present, which would reduce the inferred size of the nebula and lengthen the lifetime inferred from its proximity zone, its effects would likely be negligible in the context of \cref{fig:tqsummary}.

\subsection{\lya\ emission mechanism}

\begin{figure}
    \centering
    \includegraphics[width=\linewidth]{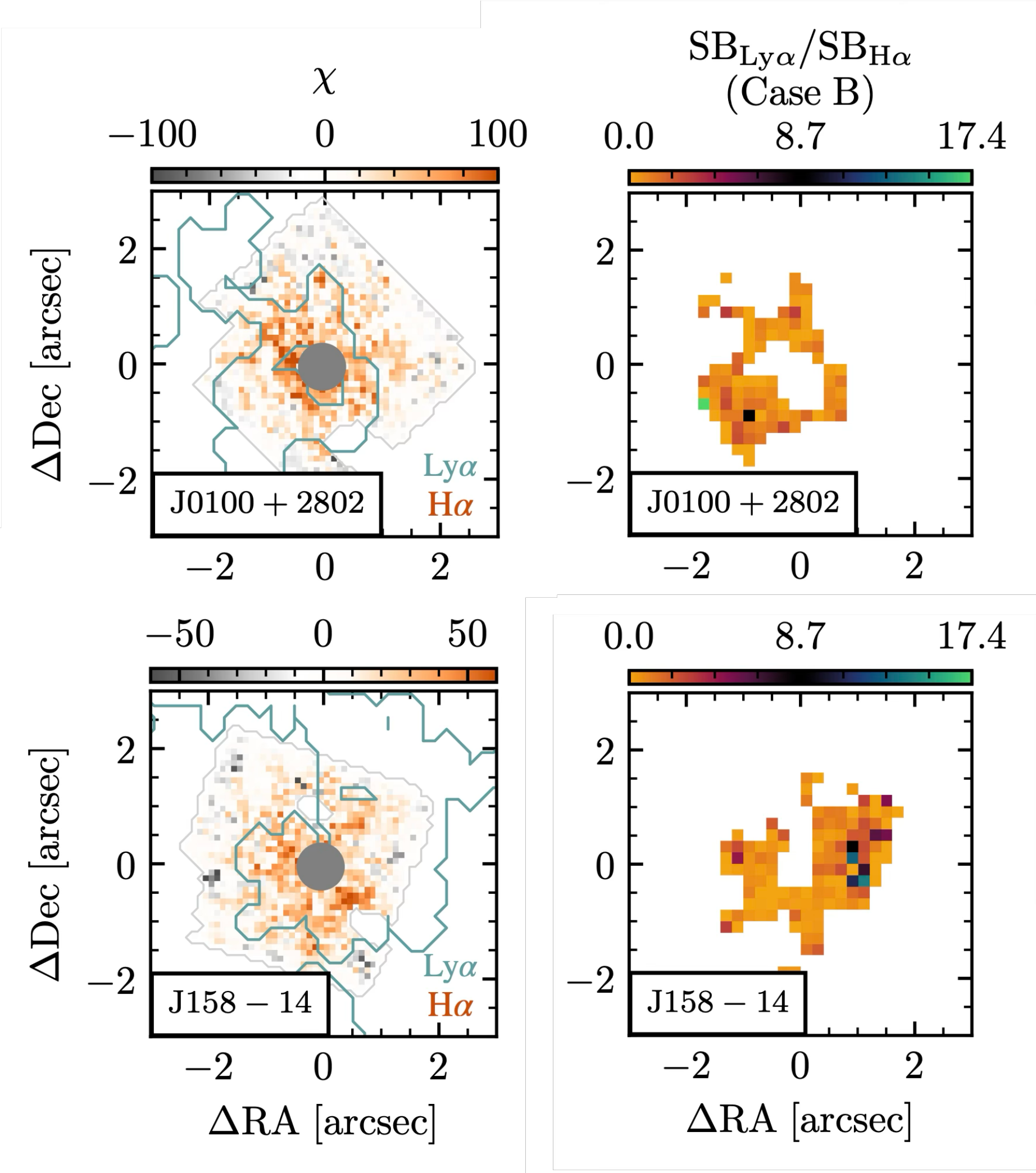}
    \caption{In the left column, we show the $\chi$ maps of the two quasars in our sample that have nebular detections in both \ha\ (this work, orange) and \lya\ \citep[blue contour][]{durovcikova_quasar_2025}. To determine the mechanism behind the \lya\ emission, we calculate \lya-to-\ha\ surface brightness ratios using the region enclosed by the \lya\ contour within the field-of-view of the NIRSpec IFU (right column). The gray circle with a radius of $0.4''$ centered on the quasar marks the region which is dominated by PSF residuals in the MUSE data (\lya) and so is excluded from the calculation.}
    \label{fig:LyaHa}
\end{figure}

Concurrent \ha\ and \lya\ nebular detections provide a unique way of addressing the uncertainty behind the emission mechanism of the \lya\ emission \cite[e.g.][]{leibler_detection_2018,langen_characterizing_2023}. Due to its complicated radiative transfer, the spatial extent of \lya\ nebulae is difficult to unambiguously interpret in the context of quasar lifetimes -- many scattering events could significantly prolong the time required to cause \lya\ emission a certain distance away from the quasar. This would render the measurements of \cite{durovcikova_quasar_2025} lower limits on the transverse quasar lifetimes. Thanks to the data set at hand, we are in a unique position to place constraints on the emission mechanism behind \lya\ nebulae at high redshift, and thus cross-validate the corresponding \lya\ nebular lifetime constraints. Two quasars in our sample -- SDSS J0100+2802 and PSO J158--14 -- have clear \lya\ and \ha\ nebular detections, which are suitable for this exercise. We display their \ha\ $\chi$ maps again in \cref{fig:LyaHa}, where we also show the contours of their corresponding \lya\ nebular detections by \cite{durovcikova_quasar_2025}.

If recombination is the dominant mechanism behind the \lya\ emission, an assumption that the lifetime measurements of \cite{durovcikova_quasar_2025} rely on, the emission of both \lya\ and \ha\ results from the hydrogen recombination cascade and is thus fixed due to selection rules governing electronic transitions in the hydrogen atom. For Case B recombination (at a temperature of $T \approx 10^4\ {\rm K}$), the expected flux ratio of \lya\ to \ha\ emission is $F_{\rm Ly\alpha}/F_{\rm H\alpha} = 8.7$ \citep{storey_recombination_1995} -- hence, $F_{\rm Ly\alpha}/F_{\rm H\alpha} \gg 8.7$ would point towards scattering playing a significant role in the interpretation of \lya\ nebular lifetimes.

In order to compute this ratio, we collapse the PSF-subtracted data cubes from both the NIRSpec IFU and MUSE in a $\pm500{\rm km/s}$ spectral window around the respective emission lines to probe a comparable physical region of emission. We then bin the NIRSpec IFU pseudo-narrowband image ($0.1''$) to the spatial resolution of MUSE ($0.2''$) and align the pixels based on the quasar location in both images. \cite{durovcikova_quasar_2025} used the region with a radius of $0.4''$ around the quasar to normalize their PSF model, and so we mask this region in our calculation (denoted by a gray circle in the left column of \cref{fig:LyaHa}). Since the MUSE data is seeing limited, we smooth the \ha\ images spatially using a Gaussian kernel with ${\rm FWHM} = 1.1''$ (for SDSS J0100+2802) and ${\rm FWHM} = 1.0''$ (for PSO J158-14), which corresponds to the average seeing during the observations. Since the pixel sizes are the same, $F_{\rm Ly\alpha}/F_{\rm H\alpha} = {\rm SB}_{\rm Ly\alpha}/{\rm SB}_{\rm H\alpha}$, and so we compute a spatially resolved map of the surface brightness ratios which we display in the right column of \cref{fig:LyaHa}. The goal here is to investigate whether the \lya\ emission detected by \cite{durovcikova_quasar_2025} is dominated by scattering or recombination -- therefore, we use all pixels enclosed by the \lya\ contours and the field-of-view of the NIRSpec IFU in \cref{fig:LyaHa} to calculate this flux ratio. This leads to $F_{\rm Ly\alpha}/F_{\rm H\alpha} = 0.6_{-0.6}^{+0.9}$ for SDSS J0100+2802, and $F_{\rm Ly\alpha}/F_{\rm H\alpha} = 0.5_{-0.5}^{+1.7}$ for PSO J158--14.

The \lya-to-\ha\ flux ratios for both quasars are very well below the expected value for Case B recombination. This means that these objects actually exhibit an overabundance of \ha\ emission compared to \lya, which supports the case of recombination-dominated \lya\ nebular emission. Dust extinction is typically invoked to explain a deficit of \lya\ photons, however, the Balmer decrements observed in these quasars do not support this as a viable explanation (Ishikawa et al. 2025, in preparation). On the other hand, attenuation by the increasingly more neutral IGM during the Epoch of Reionization is the most plausible explanation for the weak \lya\ emission compared to the \ha\ nebulae detected here, especially in light of the \lya\ damping (by $\sim15\%$ at $z=6.1$) showcased by \cite{durovcikova_chronicling_2024} that is present even close to $z=6$. Another factor affecting this ratio could be a possible contamination of the total \ha\ flux by the nearby \nii\ doublet, although this is unlikely given the narrow width of the spectral window we used to compute this ratio.

Overall, the simultaneous detections of \ha\ and \lya\ nebulae consistently point towards recombination of photoionized hydrogen gas being the dominant mechanism behind the nebular \lya\ emission in these quasars. This validates the interpretation of the recently detected \lya\ nebulae by \cite{durovcikova_quasar_2025} and paints a consistent picture of sightline-dependent obscuration being an unlikely cause of the short UV-luminous lifetimes seen at these redshifts.


\section{Implications for SMBH growth: Massive black holes, short lifetimes}\label{sec:implications}

In this work, we present deep JWST/NIRSpec IFU observations of a sample of five quasars at $z\sim6$ which exhibit small line-of-sight proximity zones, thus posing the greatest challenge to our SMBH growth models. Having access to their rest-frame optical spectral lines for the first time, we estimated their BH masses employing three different scaling relations that use their \ha\ and \hb\ emission lines, as well as a correction based on their iron emission properties. All thus derived BH masses are consistent across methods within systematic uncertainties, and, assuming these scaling relations are valid at high redshifts, they confirm the existence of billion-solar-mass SMBHs at the centers of these quasars.

We further address the challenge of growing these SMBHs in the extremely short lifetimes implied by their proximity zones ($t_{\rm Q} \lesssim 10^6\ {\rm yr}$) by searching for their rest-frame optical extended emission in \ha\ and \oiii$5007$. As proximity zones are sensitive to the escape of ionizing radiation from the quasar along our line of sight, extended nebulae provide a way of tracing the effects of quasar photoionization in the transverse direction, thereby testing for sightline-dependent obscuration effects. We find nebular detections in both \ha\ and \oiii\ in all but one quasar in our sample, and use their spatial extent to measure their nebular lifetimes in the transverse direction, expanding on the rest-UV study of \lya\ nebulae by \cite{durovcikova_quasar_2025}. We summarize our lifetime constraints in \cref{fig:tqsummary}, alongside the \lya\ nebular lifetimes from the literature, all consistently pointing towards the existence of short UV-luminous quasar lifetimes at high redshift.

Additionally, we leverage the simultaneous detections of \ha\ from this work and \lya\ from \cite{durovcikova_quasar_2025} to constrain the emission mechanism behind the \lya\ nebulae presented in their work. Our measurements make a convincing case for recombination, thereby validating the interpretation of quasar lifetimes in \cite{durovcikova_quasar_2025} and supporting the finding that sightline-dependent obscuration effects are unlikely the dominant cause of the small proximity zones.

Overall, we confirm the existence of billion-solar-mass BHs powering luminous quasars as early as less than a billion years after the Big Bang. Our results support the existence of extremely short UV-luminous quasar lifetimes, suggesting that the SMBHs hosted in these quasars have indeed been accreting for less than a million years. Even though previous episodes of UV-luminous accretion may have been present, our results confirm that these are extremely short-lived at high redshift, which makes explaining the growth of these SMBHs extremely challenging. This suggests that mechanisms such as super-Eddington accretion and/or fully dust-obscured growth may be at play -- understanding and searching for these therefore forms the next step towards understanding the assembly of the earliest SMBHs in our Universe.

\section*{Acknowledgments}

We would like to acknowledge the work of the JWST ERS TEMPLATES team, in particular Taylor Hutchison, whose jupyter notebooks provided a helpful starting point for our data reduction. Additionally, we would like to thank Emanuele Paolo Farina, Madeline Marshall and Nickolay Gnedin for helpful discussions.

This work is based on observations made with the NASA/ESA/CSA James Webb Space Telescope. The data were obtained from the Mikulski Archive for Space Telescopes at the Space Telescope Science Institute, which is operated by the Association of Universities for Research in Astronomy, Inc., under NASA contract NAS 5-03127 for JWST. These observations are associated with programs \#3079 and \#1218.

All the JWST data used in this paper can be found in MAST: \dataset[10.17909/wwb8-3421]{http://dx.doi.org/10.17909/wwb8-3421}.

M.V. gratefully acknowledges financial support from the Independent Research Fund Denmark via grant number DFF 3103-00146 and from the Carlsberg Foundation via grant CF23-0417. J.-T.S. is supported by the Deutsche Forschungsgemeinschaft (DFG, German Research
Foundation) – 518006966.

%

\vspace{5mm}




\appendix

\section{Background model for the NIRSpec IFU}\label{app:BG}

\begin{figure*}
    \centering
    \includegraphics[width=\linewidth]{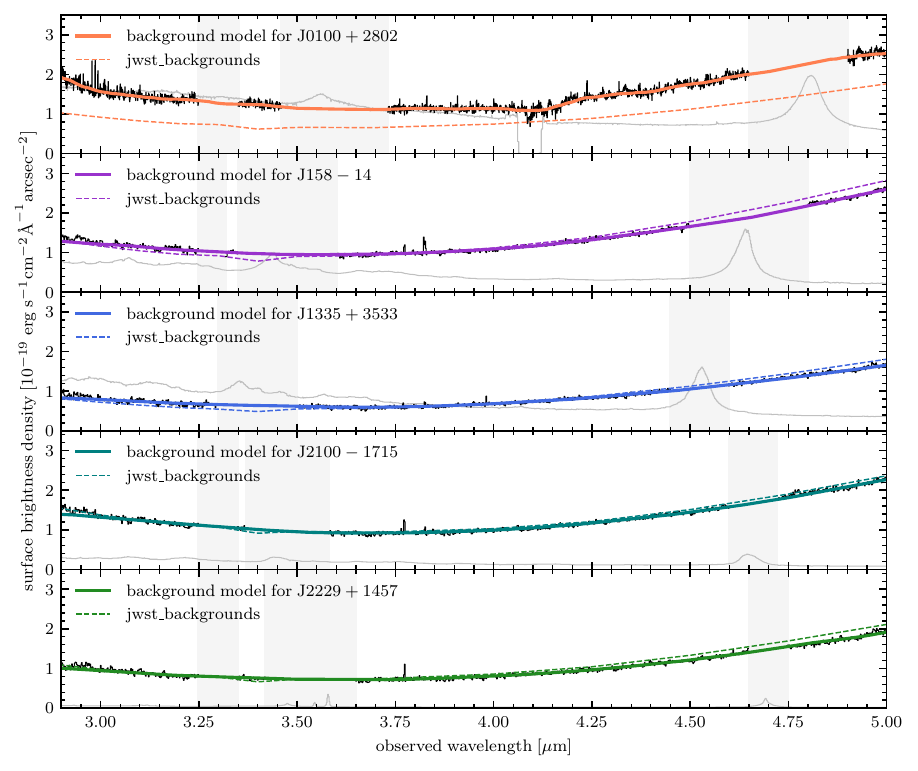}
    \caption{For each quasar, we show the extracted model for the NIRSpec IFU background. In the background of each panel, we display the mean background spectrum extracted from the data cubes as described in \S~\ref{sec:preprocessing} in black, the masked emission line wavelengths as gray shaded regions, and the quasar spectrum in gray for comparison. For comparison, we also show the predicted backgrounds from the \texttt{jwst\_backgrounds} tool (dashed lines), which match very well for the medium-resolution observations (bottom four panels), but deviate from the empirical background of the high-resolution grating target (J0100+2802).}
    \label{fig:BG}
\end{figure*}

An accurate subtraction of the sky background in the NIRSpec IFU is important for our science goal, as the NIRSpec IFU field of view, $3''\times3''$, could easily be filled up with extended emission for lifetimes of $t_{\rm Q} \lesssim 4\times10^4\ {\rm yr}$. As described in \S~\ref{sec:preprocessing}, we mask out the major emission line regions, i.e. \ha\ and \hb$+$\oiii\ (shaded gray regions in \cref{fig:BG}), and compute the mean background surface density in an annular aperture at a radius of $>1''$ away from the quasar so the surface brightness is not heavily contaminated by the quasar PSF (shown in black in all panels of \cref{fig:BG}). As the background is expected to be smoothly varying with wavelength, we interpolate and smooth the extracted background spectrum to construct the resultant background model for each data cube (shown as colored lines in \cref{fig:BG}). Note that we also mask a narrow region around $3.3\ {\rm \mu m}$, which in all cases contains a sky emission/detector artifact. In \cref{fig:BG}, we also show the background model calculated using the \texttt{jwst\_backgrounds}\footnote{\url{https://github.com/spacetelescope/jwst_backgrounds}} tool as a dashed line for comparison. This background model matches our empirical background model quite well for the quasars observed with the medium-resolution grating, however, is significantly suppressed compared to the background we observe in SDSS J0100+2802, and should therefore be generally used with caution for data analysis.

\section{Spectral regions}\label{app:specregions}

\begin{figure*}
    \centering
    \includegraphics[width=\linewidth]{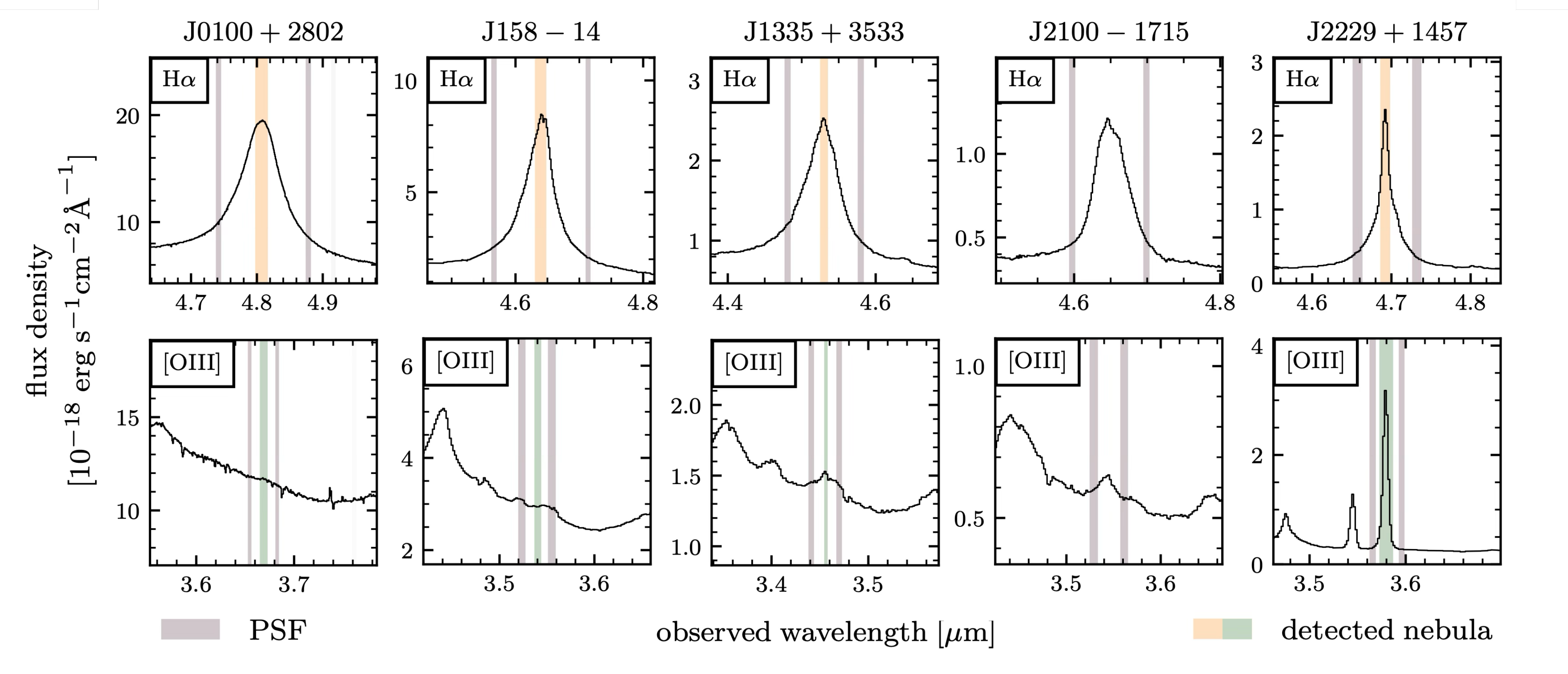}
    \caption{A summary of spectral regions used to construct the PSF (in purple) for each quasar and for both the \ha\ (top row, orange) and \oiii\ (bottom row, green) extended emission search. The spectral range of the median nebulae identified and shown in \cref{fig:nebJ0100,fig:nebJ158,fig:nebJ1335,fig:nebJ2100,fig:nebJ2229} are shown as orange/green shaded regions.}
    \label{fig:specregions}
\end{figure*}

Here we show the spectral regions that were used to construct the PSF in \S~\ref{sec:PSF}, as well as the spectral regions corresponding to the nebulae identified by the median mask as described in \S~\ref{sec:search}. The top row of \cref{fig:specregions} shows the \ha\ spectral regions (the median nebulae are highlighted in orange) and the bottom row shows the \oiii\ spectral regions (with the median nebulae highlighted in green). Note that we did not detect \ha\ or \oiii\ nebulae in the case of one quasar in our sample. We note that the velocity offset of the PSF windows ranges from $\Delta v \approx 2000 - 4500\ {\rm km/s}$ for the case of the \ha\ emission line, and $\Delta v \approx 1000\ {\rm km/s}$ for the case of \oiii. Since the nuclear \oiii\ emission is heavily blended with the iron complex and \hb\ in four out of the five quasars in our sample, we also tested, among others, an \oiii\ PSF extraction for a velocity offset of $\Delta v \approx 4000\ {\rm km/s}$ (which falls blueward of the \oiii\ doublet and redward of the prominent \feii\ emission in this spectral region) and found no appreciable difference in the resultant nebulae.

\section{Delay time to nebular emission}\label{app:delaytime}

A potential concern about using the light travel time between the quasar and the nebula to derive a constraint on the quasar lifetime is that there could be an additional delay due to line emission not following instantaneously after photoionization. This section aims to address this concern. More specifically, we estimate how long it takes for such photoionized gas to emit enough photons that an extended nebula is observable. 

Let us focus on hydrogen recombination emission (\ha), as it is also relevant to the previous nebular lifetime study that used \lya\ emission \citep{durovcikova_quasar_2025}. The Case B recombination timescale for typical nebular gas conditions is
\begin{equation}
    t_{\rm rec} = \frac{1}{n_{\rm H}\alpha_{\rm B}} \sim 10^5~{\rm yr},
\end{equation}
assuming a hydrogen number density of $n_{\rm H} \sim 1~{\rm cm^{-3}}$ \citep{cantalupo_cosmic_2014, arrigoni_battaia_deep_2015,hennawi_quasar_2015} and evaluating the recombination coefficient at a temperature of $T=10^4~{\rm K}$, which yields $\alpha_{\rm B} \approx 2.6\times10^{-13}~{\rm cm^3\ s^{-1}}$. The inverse of this timescale represents the rate of recombinations per unit volume, $R_{\rm rec} = 1/t_{\rm rec}$. 

The observed nebular emission is a result of many recombinations over a large volume, $V$. Hence, the total recombination rate in the volume relevant for nebular emission is given by
\begin{equation}
    R_{\rm rec}^{V} = R_{\rm rec}\frac{4}{3}\pi r_{\rm neb}^3 C_{\rm V},
\end{equation}
where we have approximated the nebular volume as a sphere of radius $r_{\rm neb}$. $C_{\rm V}$ is the volume-filling factor and is understood to be generally small for \lya\ emitting gas clouds, $C_{\rm V} < 10^{-5}-10^{-2}$ \citep{mccourt_characteristic_2018,pezzulli_high_2019,prochaska_quasars_2009}. The total recombination rate can be further translated into the total photon production rate via
\begin{equation}
    R_{\rm phot}^{V} = R_{\rm rec}^{V} f_{\rm phot} f_{\rm esc},
\end{equation}
where $f_{\rm phot}$ represents the fraction of \ha/\lya\ photons produced in a hydrogen recombination event, and $f_{\rm esc}$ denotes the escape fraction of those photons out of the CGM.

As an example, let us consider a 1\,${\rm arcsec^2}$ region around the quasar, which roughly corresponds to a nebular radius of $r_{\rm neb}\sim5~{\rm kpc}$ at $z=6$. This nebular size is comparable to the nebulae found in this work and is detectable with both the NIRSpec IFU as well as MUSE \citep{durovcikova_quasar_2025}. For \ha, we use $f_{\rm phot}\sim 0.4$ and $f_{\rm esc}\sim 1$. For \lya, $f_{\rm phot}\sim 0.7$ and, due to its resonant nature, we assume $f_{\rm esc}\sim 0.1$ (although this parameter is quite difficult to constrain). Assuming $C_{\rm V} = 10^{-2}$, the photon production rate in this nebular volume is $R_{\rm phot}^{V}\sim 1.6\times10^{52}~{\rm s^{-1}}$ for \ha\ and $R_{\rm phot}^{V} \sim 2.8\times10^{51}~{\rm s^{-1}}$ for \lya\ -- this extremely large number of photons produced per second is due to the large number of hydrogen atoms in the CGM undergoing recombinations (despite the per unit volume recombination timescale being as large as $t_{\rm rec}\sim 10^5~{\rm yr}$).

Given this photon production timescale, let us now estimate the time it takes to collect enough photons to reach the detection sensitivity in this work and in \cite{durovcikova_quasar_2025}. In both cases, the surface brightness sensitivity over a $1{\rm arcsec^2}$ region is on the order of ${\rm SB}\sim 10^{-18}~{\rm erg\ s^{-1}\ cm^{-2}\ arcsec^{-2}}$. Converting this limit to the luminosity at $z=6$ and for the observed emission area of $1~{\rm arcsec^2}$ yields $L = 4\pi D_{\rm L}(z)^2 {\rm SB} A_{\rm neb} \approx 4.2\times10^{41}~{\rm erg\ s^{-1}}$. Now, we can estimate the number of photons that need to be collected to reach this surface brightness sensitivity, as
\begin{equation}
    N_{\rm phot} \approx \frac{L t_{\rm exp}}{E_{\rm phot}} = \frac{L t_{\rm exp} \lambda_{\rm phot}}{hc}.
\end{equation}
Note that $\lambda_{\rm phot}$ is the rest wavelength of the emitted photon. The expected delay time between photoionization and the production of an observable nebula is then given by
\begin{equation}
    t_{\rm delay} = (1+z)\frac{N_{\rm phot}}{R_{\rm phot}^{V}} = (1+z)\frac{L t_{\rm exp} \lambda_{\rm phot}}{hcR_{\rm phot}^{V}},
\end{equation}
where the factor of $(1+z)$ accounts of time dilation.

To calculate a conservative estimate on $t_{\rm delay}$, we use the longest exposure time across the respective samples. In the case of \ha\ (this work), we use $t_{\rm exp} = 6~{\rm h}$, which yields $t_{\rm delay}^{\rm H\alpha} \approx 360~{\rm h}$ (roughly $15$ days). For \lya, $t_{\rm exp} = 8.5~{\rm h}$ \citep{durovcikova_quasar_2025} yields $t_{\rm delay}^{\rm Ly\alpha} \approx 540~{\rm h}$ (roughly $23$ days). If we assume a much smaller volume filling factor $C_{\rm V} = 10^{-5}$, we obtain $t_{\rm delay}^{\rm H\alpha} \approx 41~{\rm yr}$ and $t_{\rm delay}^{\rm Ly\alpha} \approx 62~{\rm yr}$. Even under the most conservative assumptions, the time is takes to create an observable nebula after the quasar photoionizes the CGM gas is negligible compared to the nebular lifetimes derived here and in \cite{durovcikova_quasar_2025}. Hence, the use of light travel time is justified.


\bibliography{BEES_refs}{}

\begin{thebibliography}{}
\expandafter\ifx\csname natexlab\endcsname\relax\def\natexlab#1{#1}\fi
\providecommand{\url}[1]{\href{#1}{#1}}
\providecommand{\dodoi}[1]{doi:~\href{http://doi.org/#1}{\nolinkurl{#1}}}
\providecommand{\doeprint}[1]{\href{http://ascl.net/#1}{\nolinkurl{http://ascl.net/#1}}}
\providecommand{\doarXiv}[1]{\href{https://arxiv.org/abs/#1}{\nolinkurl{https://arxiv.org/abs/#1}}}

\bibitem[{Arrigoni~Battaia {et~al.}(2015)Arrigoni~Battaia, Yang, Hennawi, Prochaska, Matsuda, Yamada, \& Hayashino}]{arrigoni_battaia_deep_2015}
Arrigoni~Battaia, F., Yang, Y., Hennawi, J.~F., {et~al.} 2015, The Astrophysical Journal, 804, 26, \dodoi{10.1088/0004-637X/804/1/26}

\bibitem[{Arrigoni Battaia {et~al.}(2019)Arrigoni Battaia, Hennawi, Prochaska, Oñorbe, Farina, Cantalupo, \& Lusso}]{arrigonibattaia_qso_2019}
Arrigoni Battaia, F., Hennawi, J.~F., Prochaska, J.~X., {et~al.} 2019, Monthly Notices of the Royal Astronomical Society, 482, 3162, \dodoi{10.1093/mnras/sty2827}

\bibitem[{Bacon {et~al.}(2010)Bacon, Accardo, Adjali, Anwand, Bauer, Biswas, Blaizot, Boudon, Brau-Nogue, Brinchmann, Caillier, Capoani, Carollo, Contini, Couderc, Daguisé, Deiries, Delabre, Dreizler, Dubois, Dupieux, Dupuy, Emsellem, Fechner, Fleischmann, François, Gallou, Gharsa, Glindemann, Gojak, Guiderdoni, Hansali, Hahn, Jarno, Kelz, Koehler, Kosmalski, Laurent, Le~Floch, Lilly, Lizon, Loupias, Manescau, Monstein, Nicklas, Olaya, Pares, Pasquini, Pécontal-Rousset, Pelló, Petit, Popow, Reiss, Remillieux, Renault, Roth, Rupprecht, Serre, Schaye, Soucail, Steinmetz, Streicher, Stuik, Valentin, Vernet, Weilbacher, Wisotzki, \& Yerle}]{bacon_muse_2010}
Bacon, R., Accardo, M., Adjali, L., {et~al.} 2010, 7735, 773508, \dodoi{10.1117/12.856027}

\bibitem[{Bolgar {et~al.}(2018)Bolgar, Eames, Hottier, \& Semelin}]{bolgar_imprints_2018}
Bolgar, F., Eames, E., Hottier, C., \& Semelin, B. 2018, Monthly Notices of the Royal Astronomical Society, 478, 5564, \dodoi{10.1093/mnras/sty1293}

\bibitem[{Borisova {et~al.}(2016{\natexlab{a}})Borisova, Lilly, Cantalupo, Prochaska, Rakic, \& Worseck}]{borisova_constraining_2016}
Borisova, E., Lilly, S.~J., Cantalupo, S., {et~al.} 2016{\natexlab{a}}, The Astrophysical Journal, 830, 120, \dodoi{10.3847/0004-637X/830/2/120}

\bibitem[{Borisova {et~al.}(2016{\natexlab{b}})Borisova, Cantalupo, Lilly, Marino, Gallego, Bacon, Blaizot, Bouché, Brinchmann, Carollo, Caruana, Finley, Herenz, Richard, Schaye, Straka, Turner, Urrutia, Verhamme, \& Wisotzki}]{borisova_ubiquitous_2016}
Borisova, E., Cantalupo, S., Lilly, S.~J., {et~al.} 2016{\natexlab{b}}, The Astrophysical Journal, 831, 39, \dodoi{10.3847/0004-637X/831/1/39}

\bibitem[{Cantalupo {et~al.}(2014)Cantalupo, Arrigoni-Battaia, Prochaska, Hennawi, \& Madau}]{cantalupo_cosmic_2014}
Cantalupo, S., Arrigoni-Battaia, F., Prochaska, J.~X., Hennawi, J.~F., \& Madau, P. 2014, Nature, 506, 63, \dodoi{10.1038/nature12898}

\bibitem[{Costa {et~al.}(2022)Costa, Arrigoni~Battaia, Farina, Keating, Rosdahl, \& Kimm}]{costa_agn-driven_2022}
Costa, T., Arrigoni~Battaia, F., Farina, E.~P., {et~al.} 2022, Monthly Notices of the Royal Astronomical Society, 517, 1767, \dodoi{10.1093/mnras/stac2432}

\bibitem[{Davies {et~al.}(2019)Davies, Hennawi, \& Eilers}]{davies_evidence_2019}
Davies, F.~B., Hennawi, J.~F., \& Eilers, A.-C. 2019, The Astrophysical Journal Letters, 884, L19, \dodoi{10.3847/2041-8213/ab42e3}

\bibitem[{Davies {et~al.}(2020)Davies, Wang, Eilers, \& Hennawi}]{davies_constraining_2020}
Davies, F.~B., Wang, F., Eilers, A.-C., \& Hennawi, J.~F. 2020, The Astrophysical Journal Letters, 904, L32, \dodoi{10.3847/2041-8213/abc61f}

\bibitem[{Doj\v{c}inovi\'{c} {et~al.}(2023)Doj\v{c}inovi\'{c}, Kova\v{c}evi\'{c}-Doj\v{c}inovi\'{c}, \& Popovi\'{c}}]{dojcinovic_flux_2023}
Doj\v{c}inovi\'{c}, I., Kova\v{c}evi\'{c}-Doj\v{c}inovi\'{c}, J., \& Popovi\'{c}, L. 2023, Advances in Space Research, 71, 1219, \dodoi{10.1016/j.asr.2022.04.041}

\bibitem[{Drake {et~al.}(2019)Drake, Farina, Neeleman, Walter, Venemans, Banados, Mazzucchelli, \& Decarli}]{drake_ly_2019}
Drake, A.~B., Farina, E.~P., Neeleman, M., {et~al.} 2019, The Astrophysical Journal, 881, 131, \dodoi{10.3847/1538-4357/ab2984}

\bibitem[{Du \& Wang(2019)}]{du_radiusluminosity_2019}
Du, P., \& Wang, J.-M. 2019, The Astrophysical Journal, 886, 42, \dodoi{10.3847/1538-4357/ab4908}

\bibitem[{Eilers {et~al.}(2018)Eilers, Hennawi, \& Davies}]{eilers_first_2018}
Eilers, A.-C., Hennawi, J.~F., \& Davies, F.~B. 2018, The Astrophysical Journal, 867, 30, \dodoi{10.3847/1538-4357/aae081}

\bibitem[{Eilers {et~al.}(2021)Eilers, Hennawi, Davies, \& Simcoe}]{eilers_detecting_2021}
Eilers, A.-C., Hennawi, J.~F., Davies, F.~B., \& Simcoe, R.~A. 2021, The Astrophysical Journal, 917, 38, \dodoi{10.3847/1538-4357/ac0a76}

\bibitem[{Eilers {et~al.}(2020)Eilers, Hennawi, Decarli, Davies, Venemans, Walter, Bañados, Fan, Farina, Mazzucchelli, Novak, Schindler, Simcoe, Wang, \& Yang}]{eilers_detecting_2020}
Eilers, A.-C., Hennawi, J.~F., Decarli, R., {et~al.} 2020, The Astrophysical Journal, 900, 37, \dodoi{10.3847/1538-4357/aba52e}

\bibitem[{Eilers {et~al.}(2023)Eilers, Simcoe, Yue, Mackenzie, Matthee, Ďurovčíková, Kashino, Bordoloi, \& Lilly}]{eilers_eiger_2023}
Eilers, A.-C., Simcoe, R.~A., Yue, M., {et~al.} 2023, The Astrophysical Journal, 950, 68, \dodoi{10.3847/1538-4357/acd776}

\bibitem[{Eilers {et~al.}(2024)Eilers, Mackenzie, Pizzati, Matthee, Hennawi, Zhang, Bordoloi, Kashino, Lilly, Naidu, Simcoe, Yue, Frenk, Helly, Schaller, \& Schaye}]{eilers_eiger_2024}
Eilers, A.-C., Mackenzie, R., Pizzati, E., {et~al.} 2024, The Astrophysical Journal, 974, 275, \dodoi{10.3847/1538-4357/ad778b}

\bibitem[{Eilers {et~al.}(2025)Eilers, Yue, Matthee, Hennawi, Davies, Simcoe, Teague, Bordoloi, Brammer, Kang, Kashino, Mackenzie, Naidu, \& Navarrete}]{eilers_light_2025}
Eilers, A.-C., Yue, M., Matthee, J., {et~al.} 2025, The {Light} {Echo} of a {High}-{Redshift} {Quasar} mapped with {Lyman}-\$$\alpha$\$ {Tomography},  arXiv, \dodoi{10.48550/arXiv.2509.05417}

\bibitem[{Fan {et~al.}(2023)Fan, Bañados, \& Simcoe}]{fan_quasars_2023}
Fan, X., Bañados, E., \& Simcoe, R.~A. 2023, Annual Review of Astronomy and Astrophysics, 61, 373, \dodoi{10.1146/annurev-astro-052920-102455}

\bibitem[{Fan {et~al.}(2006)Fan, Strauss, Becker, White, Gunn, Knapp, Richards, Schneider, Brinkmann, \& Fukugita}]{fan_constraining_2006}
Fan, X., Strauss, M.~A., Becker, R.~H., {et~al.} 2006, The Astronomical Journal, 132, 117, \dodoi{10.1086/504836}

\bibitem[{Farina {et~al.}(2017)Farina, Venemans, Decarli, Hennawi, Walter, Bañados, Mazzucchelli, Cantalupo, Arrigoni-Battaia, \& McGreer}]{farina_mapping_2017}
Farina, E.~P., Venemans, B.~P., Decarli, R., {et~al.} 2017, The Astrophysical Journal, 848, 78, \dodoi{10.3847/1538-4357/aa8df4}

\bibitem[{Farina {et~al.}(2019)Farina, Arrigoni-Battaia, Costa, Walter, Hennawi, Drake, Decarli, Gutcke, Mazzucchelli, Neeleman, Georgiev, Eilers, Davies, Bañados, Fan, Onoue, Schindler, Venemans, Wang, Yang, Rabien, \& Busoni}]{farina_requiem_2019}
Farina, E.~P., Arrigoni-Battaia, F., Costa, T., {et~al.} 2019, The Astrophysical Journal, 887, 196, \dodoi{10.3847/1538-4357/ab5847}

\bibitem[{Foreman-Mackey {et~al.}(2013)Foreman-Mackey, Hogg, Lang, \& Goodman}]{foreman-mackey_emcee_2013}
Foreman-Mackey, D., Hogg, D.~W., Lang, D., \& Goodman, J. 2013, Publications of the Astronomical Society of the Pacific, 125, 306, \dodoi{10.1086/670067}

\bibitem[{Greene \& Ho(2005)}]{greene_estimating_2005}
Greene, J.~E., \& Ho, L.~C. 2005, The Astrophysical Journal, 630, 122, \dodoi{10.1086/431897}

\bibitem[{Hennawi \& Prochaska(2013)}]{hennawi_quasars_2013}
Hennawi, J.~F., \& Prochaska, J.~X. 2013, The Astrophysical Journal, 766, 58, \dodoi{10.1088/0004-637X/766/1/58}

\bibitem[{Hennawi {et~al.}(2015)Hennawi, Prochaska, Cantalupo, \& Arrigoni-Battaia}]{hennawi_quasar_2015}
Hennawi, J.~F., Prochaska, J.~X., Cantalupo, S., \& Arrigoni-Battaia, F. 2015, Science, 348, 779, \dodoi{10.1126/science.aaa5397}

\bibitem[{Inayoshi {et~al.}(2020)Inayoshi, Visbal, \& Haiman}]{inayoshi_assembly_2020}
Inayoshi, K., Visbal, E., \& Haiman, Z. 2020, Annual Review of Astronomy and Astrophysics, 58, 27, \dodoi{10.1146/annurev-astro-120419-014455}

\bibitem[{Langen {et~al.}(2023)Langen, Cantalupo, Steidel, Chen, Pezzulli, \& Gallego}]{langen_characterizing_2023}
Langen, V., Cantalupo, S., Steidel, C.~C., {et~al.} 2023, Monthly Notices of the Royal Astronomical Society, 519, 5099, \dodoi{10.1093/mnras/stac3205}

\bibitem[{Leibler {et~al.}(2018)Leibler, Cantalupo, Holden, \& Madau}]{leibler_detection_2018}
Leibler, C.~N., Cantalupo, S., Holden, B.~P., \& Madau, P. 2018, Monthly Notices of the Royal Astronomical Society, 480, 2094, \dodoi{10.1093/mnras/sty1764}

\bibitem[{Liu {et~al.}(2013)Liu, Zakamska, Greene, Nesvadba, \& Liu}]{liu_observations_2013}
Liu, G., Zakamska, N.~L., Greene, J.~E., Nesvadba, N. P.~H., \& Liu, X. 2013, Monthly Notices of the Royal Astronomical Society, 430, 2327, \dodoi{10.1093/mnras/stt051}

\bibitem[{McCourt {et~al.}(2018)McCourt, Oh, O'Leary, \& Madigan}]{mccourt_characteristic_2018}
McCourt, M., Oh, S.~P., O'Leary, R., \& Madigan, A.-M. 2018, Monthly Notices of the Royal Astronomical Society, 473, 5407, \dodoi{10.1093/mnras/stx2687}

\bibitem[{Morey {et~al.}(2021)Morey, Eilers, Davies, Hennawi, \& Simcoe}]{morey_estimating_2021}
Morey, K.~A., Eilers, A.-C., Davies, F.~B., Hennawi, J.~F., \& Simcoe, R.~A. 2021, The Astrophysical Journal, 921, 88, \dodoi{10.3847/1538-4357/ac1c70}

\bibitem[{Osterbrock \& Ferland(2006)}]{osterbrock_astrophysics_2006}
Osterbrock, D.~E., \& Ferland, G.~J. 2006, Astrophysics of gaseous nebulae and active galactic nuclei.
\newblock \url{https://ui.adsabs.harvard.edu/abs/2006agna.book.....O/abstract}

\bibitem[{Peng {et~al.}(2025)Peng, Arrigoni~Battaia, Vishwas, Li, Iani, Sun, Li, Ferkinhoff, Stacey, Cai, \& Ivison}]{peng_direct_2025}
Peng, B., Arrigoni~Battaia, F., Vishwas, A., {et~al.} 2025, Astronomy \&amp; Astrophysics, Volume 694, id.L1, 12 pp., 694, L1, \dodoi{10.1051/0004-6361/202452610}

\bibitem[{Pezzulli \& Cantalupo(2019)}]{pezzulli_high_2019}
Pezzulli, G., \& Cantalupo, S. 2019, Monthly Notices of the Royal Astronomical Society, 486, 1489, \dodoi{10.1093/mnras/stz906}

\bibitem[{{Planck Collaboration} {et~al.}(2020){Planck Collaboration}, Aghanim, Akrami, Ashdown, Aumont, Baccigalupi, Ballardini, Banday, Barreiro, Bartolo, Basak, Battye, Benabed, Bernard, Bersanelli, Bielewicz, Bock, Bond, Borrill, Bouchet, Boulanger, Bucher, Burigana, Butler, Calabrese, Cardoso, Carron, Challinor, Chiang, Chluba, Colombo, Combet, Contreras, Crill, Cuttaia, De~Bernardis, De~Zotti, Delabrouille, Delouis, Di~Valentino, Diego, Doré, Douspis, Ducout, Dupac, Dusini, Efstathiou, Elsner, Enßlin, Eriksen, Fantaye, Farhang, Fergusson, Fernandez-Cobos, Finelli, Forastieri, Frailis, Fraisse, Franceschi, Frolov, Galeotta, Galli, Ganga, Génova-Santos, Gerbino, Ghosh, González-Nuevo, Górski, Gratton, Gruppuso, Gudmundsson, Hamann, Handley, Hansen, Herranz, Hildebrandt, Hivon, Huang, Jaffe, Jones, Karakci, Keihänen, Keskitalo, Kiiveri, Kim, Kisner, Knox, Krachmalnicoff, Kunz, Kurki-Suonio, Lagache, Lamarre, Lasenby, Lattanzi, Lawrence, Le~Jeune, Lemos, Lesgourgues, Levrier, Lewis, Liguori, Lilje,
  Lilley, Lindholm, López-Caniego, Lubin, Ma, Macías-Pérez, Maggio, Maino, Mandolesi, Mangilli, Marcos-Caballero, Maris, Martin, Martinelli, Martínez-González, Matarrese, Mauri, McEwen, Meinhold, Melchiorri, Mennella, Migliaccio, Millea, Mitra, Miville-Deschênes, Molinari, Montier, Morgante, Moss, Natoli, Nørgaard-Nielsen, Pagano, Paoletti, Partridge, Patanchon, Peiris, Perrotta, Pettorino, Piacentini, Polastri, Polenta, Puget, Rachen, Reinecke, Remazeilles, Renzi, Rocha, Rosset, Roudier, Rubiño-Martín, Ruiz-Granados, Salvati, Sandri, Savelainen, Scott, Shellard, Sirignano, Sirri, Spencer, Sunyaev, Suur-Uski, Tauber, Tavagnacco, Tenti, Toffolatti, Tomasi, Trombetti, Valenziano, Valiviita, Van~Tent, Vibert, Vielva, Villa, Vittorio, Wandelt, Wehus, White, White, Zacchei, \& Zonca}]{planck_collaboration_planck_2020}
{Planck Collaboration}, Aghanim, N., Akrami, Y., {et~al.} 2020, Astronomy \& Astrophysics, 641, A6, \dodoi{10.1051/0004-6361/201833910}

\bibitem[{Prochaska \& Hennawi(2009)}]{prochaska_quasars_2009}
Prochaska, J.~X., \& Hennawi, J.~F. 2009, The Astrophysical Journal, 690, 1558, \dodoi{10.1088/0004-637X/690/2/1558}

\bibitem[{Rauscher(2024)}]{rauscher_nsclean_2024}
Rauscher, B.~J. 2024, Publications of the Astronomical Society of the Pacific, 136, 015001, \dodoi{10.1088/1538-3873/ad1b36}

\bibitem[{Richards {et~al.}(2006)Richards, Lacy, Storrie-Lombardi, Hall, Gallagher, Hines, Fan, Papovich, Vanden~Berk, Trammell, Schneider, Vestergaard, York, Jester, Anderson, Budavári, \& Szalay}]{Richards_2006}
Richards, G.~T., Lacy, M., Storrie-Lombardi, L.~J., {et~al.} 2006, The Astrophysical Journal Supplement Series, 166, 470, \dodoi{10.1086/506525}

\bibitem[{Satyavolu {et~al.}(2023)Satyavolu, Kulkarni, Keating, \& Haehnelt}]{satyavolu_need_2023}
Satyavolu, S., Kulkarni, G., Keating, L.~C., \& Haehnelt, M.~G. 2023, Monthly Notices of the Royal Astronomical Society, 521, 3108, \dodoi{10.1093/mnras/stad729}

\bibitem[{Schindler {et~al.}(2020)Schindler, Farina, Bañados, Eilers, Hennawi, Onoue, Venemans, Walter, Wang, Davies, Decarli, Rosa, Drake, Fan, Mazzucchelli, Rix, Worseck, \& Yang}]{schindler_x-shooteralma_2020}
Schindler, J.-T., Farina, E.~P., Bañados, E., {et~al.} 2020, The Astrophysical Journal, 905, 51, \dodoi{10.3847/1538-4357/abc2d7}

\bibitem[{Shen \& Ho(2014)}]{shen_diversity_2014}
Shen, Y., \& Ho, L.~C. 2014, Nature, 513, 210, \dodoi{10.1038/nature13712}

\bibitem[{Shen \& Liu(2012)}]{shen_comparing_2012}
Shen, Y., \& Liu, X. 2012, The Astrophysical Journal, 753, 125, \dodoi{10.1088/0004-637X/753/2/125}

\bibitem[{Shen {et~al.}(2011)Shen, Richards, Strauss, Hall, Schneider, Snedden, Bizyaev, Brewington, Malanushenko, Malanushenko, Oravetz, Pan, \& Simmons}]{shen_catalog_2011}
Shen, Y., Richards, G.~T., Strauss, M.~A., {et~al.} 2011, The Astrophysical Journal Supplement Series, 194, 45, \dodoi{10.1088/0067-0049/194/2/45}

\bibitem[{Storey \& Hummer(1995)}]{storey_recombination_1995}
Storey, P.~J., \& Hummer, D.~G. 1995, Monthly Notices of the Royal Astronomical Society, 272, 41, \dodoi{10.1093/mnras/272.1.41}

\bibitem[{Sulentic {et~al.}(2000)Sulentic, Zwitter, Marziani, \& Dultzin-Hacyan}]{sulentic_eigenvector_2000}
Sulentic, J.~W., Zwitter, T., Marziani, P., \& Dultzin-Hacyan, D. 2000, The Astrophysical Journal, Volume 536, Issue 1, pp. L5-L9., 536, L5, \dodoi{10.1086/312717}

\bibitem[{Trainor \& Steidel(2013)}]{trainor_constraints_2013}
Trainor, R., \& Steidel, C.~C. 2013, The Astrophysical Journal Letters, 775, L3, \dodoi{10.1088/2041-8205/775/1/L3}

\bibitem[{Trebitsch {et~al.}(2019)Trebitsch, Volonteri, \& Dubois}]{trebitsch_black_2019}
Trebitsch, M., Volonteri, M., \& Dubois, Y. 2019, Monthly Notices of the Royal Astronomical Society, 487, 819, \dodoi{10.1093/mnras/stz1280}

\bibitem[{Vestergaard \& Osmer(2009)}]{vestergaard_mass_2009}
Vestergaard, M., \& Osmer, P.~S. 2009, The Astrophysical Journal, 699, 800, \dodoi{10.1088/0004-637X/699/1/800}

\bibitem[{Vestergaard \& Peterson(2006)}]{vestergaard_determining_2006}
Vestergaard, M., \& Peterson, B.~M. 2006, The Astrophysical Journal, 641, 689, \dodoi{10.1086/500572}

\bibitem[{Véron-Cetty {et~al.}(2004)Véron-Cetty, Joly, \& Véron}]{veron-cetty_unusual_2004}
Véron-Cetty, M.~P., Joly, M., \& Véron, P. 2004, Astronomy and Astrophysics, 417, 515, \dodoi{10.1051/0004-6361:20035714}

\bibitem[{Yang {et~al.}(2023)Yang, Wang, Fan, Hennawi, Barth, Bañados, Sun, Liu, Cai, Jiang, Li, Onoue, Schindler, Shen, Wu, Bhowmick, Bieri, Blecha, Bosman, Champagne, Colina, Connor, Costa, Davies, Decarli, De~Rosa, Drake, Egami, Eilers, Evans, Farina, Habouzit, Haiman, Jin, Jun, Kakiichi, Khusanova, Kulkarni, Loiacono, Lupi, Mazzucchelli, Pan, Rojas-Ruiz, Strauss, Tee, Trakhtenbrot, Trebitsch, Venemans, Vestergaard, Volonteri, Walter, Xie, Yue, Zhang, Zhang, \& Zou}]{yang_spectroscopic_2023}
Yang, J., Wang, F., Fan, X., {et~al.} 2023, The Astrophysical Journal, 951, L5, \dodoi{10.3847/2041-8213/acc9c8}

\bibitem[{Yue {et~al.}(2023)Yue, Eilers, Simcoe, Belli, Davies, DePalma, Hennawi, Mason, Muñoz, Nelson, \& Tacchella}]{yue_detecting_2023}
Yue, M., Eilers, A.-C., Simcoe, R.~A., {et~al.} 2023, The Astrophysical Journal, 950, 105, \dodoi{10.3847/1538-4357/accf20}

\bibitem[{Yue {et~al.}(2024)Yue, Eilers, Simcoe, Mackenzie, Matthee, Kashino, Bordoloi, Lilly, \& Naidu}]{yue_eiger_2024}
---. 2024, The Astrophysical Journal, 966, 176, \dodoi{10.3847/1538-4357/ad3914}

\bibitem[{Ďurovčíková {et~al.}(2024)Ďurovčíková, Eilers, Chen, Satyavolu, Kulkarni, Simcoe, Keating, Haehnelt, \& Bañados}]{durovcikova_chronicling_2024}
Ďurovčíková, D., Eilers, A.-C., Chen, H., {et~al.} 2024, The Astrophysical Journal, Volume 969, Issue 2, id.162, 22 pp., 969, 162, \dodoi{10.3847/1538-4357/ad4888}

\bibitem[{Ďurovčíková {et~al.}(2025{\natexlab{a}})Ďurovčíková, Eilers, Meyer, Farina, Bañados, Davies, Hennawi, Mazzucchelli, Simcoe, \& Walter}]{durovcikova_quasar_2025}
Ďurovčíková, D., Eilers, A.-C., Meyer, R.~A., {et~al.} 2025{\natexlab{a}}, eprint arXiv:2505.00080, arXiv:2505.00080, \dodoi{10.48550/arXiv.2505.00080}

\bibitem[{Ďurovčíková {et~al.}(2025{\natexlab{b}})Ďurovčíková, Eilers, Simcoe, Welsh, Meyer, Matthee, Ryan-Weber, Yue, Katz, Satyavolu, Becker, Davies, \& Farina}]{durovcikova_extremely_2025}
Ďurovčíková, D., Eilers, A.-C., Simcoe, R.~A., {et~al.} 2025{\natexlab{b}}, The Astrophysical Journal Letters, Volume 987, Issue 2, id.L33, 7 pp., 987, L33, \dodoi{10.3847/2041-8213/ade71c}

\end{thebibliography}
\bibliographystyle{aasjournal}



\end{document}